\documentclass[12pt]{article}
\usepackage{amsmath,amssymb}
\usepackage{bm}

\catcode`\@=11

\ifx\documentclass\undefined
 \def\@sect#1#2#3#4#5#6[#7]#8{\ifnum #2>\c@secnumdepth
     \let\@svsec\@empty\else
     \refstepcounter{#1}\edef\@svsec{\csname prefix#1\endcsname
        \csname the#1\endcsname\hskip 1em}\fi
     \@tempskipa #5\relax
      \ifdim \@tempskipa>\z@
        \begingroup #6\relax
          \@hangfrom{\hskip #3\relax\@svsec}{\interlinepenalty \@M #8\par}%
        \endgroup
       \csname #1mark\endcsname{#7}\addcontentsline
         {toc}{#1}{\ifnum #2>\c@secnumdepth \else
                      \protect\numberline{\csname the#1\endcsname}\fi
                    #7}\else
        \def\@svsechd{#6\hskip #3\relax  %% \relax added 2 May 90
                   \@svsec #8\csname #1mark\endcsname
                      {#7}\addcontentsline
                           {toc}{#1}{\ifnum #2>\c@secnumdepth \else
                             \protect\numberline{\csname the#1\endcsname}\fi
                       #7}}\fi
     \@xsect{#5}}
\else
    \def\@seccntformat#1{\csname prefix#1\endcsname
        \csname the#1\endcsname\quad}
\fi

\@addtoreset{equation}{section}   % Makes \section reset 'equation' counter.
\def\theequation{\arabic{section}.\arabic{equation}}

\def\thebibliography#1{\section*{References\@mkboth
 {REFERENCES}{REFERENCES}}\list
 {\leftbibmark\arabic{enumi}\rightbibmark}{
 \settowidth\labelwidth{\leftbibmark #1\rightbibmark}\leftmargin\labelwidth
 \advance\leftmargin\labelsep
 \usecounter{enumi}}
 \def\newblock{\hskip .11em plus .33em minus -.07em}
 \sloppy\clubpenalty4000\widowpenalty4000
 \sfcode`\.=1000\relax}

\def\@citex[#1]#2{\if@filesw\immediate\write\@auxout{\string\citation{#2}}\fi
  \def\@citea{}\@cite{\@for\@citeb:=#2\do
    {\@citea\def\@citea{,\penalty\@m\ }\@ifundefined
       {b@\@citeb}{{\bf ?}\@warning
       {Citation `\@citeb' on page \thepage \space undefined}}%
\hbox{\csname b@\@citeb\endcsname\citemarkdelim}}}{#1}}

\def\@cite#1#2{\leftcitemark{#1 \if@tempswa , #2\fi}\rightcitemark}
\def\leftcitemark{[}
\def\rightcitemark{]}
\def\citemarkdelim{}
\def\leftbibmark{[}
\def\rightbibmark{]}

\catcode`\@=12

\usepackage{amsmath}
\usepackage{amssymb}

\begin{document}

\begin{titlepage}

\vspace{2cm}

\begin{center}\large\bf
Construction and applications of\\ 
the manifestly gauge invariant expressions of\\ 
the solutions of the cosmological perturbation theory\\
\end{center}

\begin{center}
Takashi Hamazaki\footnote{email address: yj4t-hmzk@asahi-net.or.jp}
\end{center}

\begin{center}
%\begin{center}
Kamiyugi 3-3-4-606 Hachioji-city\\
Tokyo 192-0373 Japan\\
\end{center}

%T2>Abstract
\begin{center}\bf Abstract\end{center}

After giving how to construct the gauge invariant perturbation variable in an arbitrarily high
order and in the arbitrary background spacetime,
we consider the manifestly gauge invariant theory of the cosmological perturbations 
in the long wavelength limit with the spatially
flat homogeneous isotropic universe being the background spacetime.
In the previous paper, the physical laws, such as the evolution equations,
the constraint equations and the junction conditions become manifestly gauge invariant,
by writing them in terms of only
the single time background/scalar like objects defined by our previous paper.
In the present paper, by extending our definition of the background/scalar like objects
from the single time case to the many time case and by writing the solution of the physical law
in the form where many time background/scalar like objects are vanishing,
the solutions become the manifestly gauge invariant.
We derive the formula changing the bases of the many time background/scalar like 
objects by which we can change the time slices for many times appearing in the solution 
including the initial time and the final time.
We use this formula to our treatment of the evolution 
of the several slow rolling scalar fields using the $\tau$ function introduced 
by our previous paper.
In the manifestly gauge invariant manner, we discuss the solution
of the many step reheating, that is, the reheating with the arbitrarily many energy transfer steps. 
Using the energy density $\rho$ as the evolution parameter, we discuss
how well the junction model in which the energy transfers from the oscillatory scalar field fluid 
to the radiation fluid are described by the metric junctions approximates
the reheating process described by the differential equations 
with the decay terms.
By using the useful parametrization of the many step reheating,
we manifestly prove that in the many step reheating system where any initial perturbation of 
each component does not become extremely large compared with the initial perturbations of the other components,
compared with the effects of the prominent fluctuation generation processes the nearest to the present time 
the effects of all the fluctuations generated by the previous fluctuation generation processes become negligibly small.
Therefore it can be concluded that in order to know the first/second order Bardeen 
parameters at the present time it is sufficient to calculate only the effects of the 
fluctuation of the scalar field fluid which becomes energetically dominant lastly 
(the curvaton mechanism) and the fluctuation of the decay parameter by which 
this scalar field fluid decays into the radiation (the modulated reheating mechanism) .

\vspace{1cm}

  
PACS number(s):98.80.Cq

\end{titlepage}

\section{Introduction}

In the inflation paradigm \cite{Sato1981} \cite{Guth.1981}, after being stretched by 
the inflationary expansion, the cosmological perturbations generated by 
the slow rolling scalar fields and causing the structure formation and 
the fluctuations of the cosmic microwave background radiation stay on the superhorizon
scales until they enter into the horizon in the post Friedmann epoch
\cite{Kodama1984} \cite{Mukhanov1992} \cite{Kodama1996} \cite{Hamazaki1996}
\cite{Gordon2001}.
Therefore we want to calculate the evolutions of the cosmological perturbations
on superhorizon scales including nonlinear levels in the case where the background spacetime is 
the spatially flat homogeneous isotropic universe.
It was established that the evolutions of the cosmological perturbations in the long 
wavelength limit are constructed from the derivatives with respect to the solution constants of 
the solutions of the corresponding exactly homogeneous universe 
\cite{Taruya1998} \cite{Kodama1998}.
This fact has been used and has been developed in the various situations, by the author of the 
present paper
\cite{Hamazaki2002} \cite{Hamazaki2004} \cite{Hamazaki2008} \cite{Hamazaki2008.2} 
\cite{Hamazaki2011} and with the reinterpretations in the contexts of each authors' researches 
\cite{Sasaki1998} \cite{Lyth2005},  \cite{Wands2000}.
In the most general situation in which plural scalar fields and plural perfect fluids coexist, 
the solution of the evolutions of the cosmological perturbations in the long wavelength limit 
based on the corresponding exactly homogeneous perturbations are manifestly presented in the 
paper \cite{Hamazaki2008} and the paper \cite{Hamazaki2008.2} in the linear level and 
in the nonlinear level, respectively.
Such approach was pioneered by the paper \cite{Polarski1992} where in the linear perturbation 
the expression of the adiabatic growing mode was given in terms of the exactly homogeneous 
solution, whose existence reason was given in the paper \cite{Kodama1998} and where the authors 
tried to give the expressions of the other perturbation solutions in term of the background variables
using the fact that all the coefficients of the linear perturbation equation with vanishing wavenumber
are written in term of the background variables only.

The perturbations are the differences between the physical quantities of the background 
homogeneous isotropic universe and those of the perturbed real universe.
How to connect the physical quantity of the perturbed universe with the spacetime point 
of the corresponding homogeneous universe in order to take the difference is called the gauge.
The changes of the perturbations generated by changing the gauge are apparent, do not have
any physical meanings.
The physical perturbations must be described by the perturbation variables       
which are gauge invariant, that is, do not change by changing the gauge.
The theories in which the physical laws are described by the gauge invarinat perturbation 
variables only are called the gauge invariant perturbation theories. 
The gauge invariant theory of the cosmological perturbation was constructed in the papers
\cite{Bardeen.J1980} \cite{Kodama1984} \cite{Kodama1987} \cite{Mukhanov1992} in the 
linear perturbation level, and in the papers \cite{Malik2009} \cite{Nakamura2006} 
in the second perturbation level.
In an arbitrarily higher order nonlinear perturbation level, 
the treatment of the cosmological perturbations on the superhorizon scales 
in the manifestly gauge invariant manner was attained in the papers 
\cite{Hamazaki2008.2} \cite{Hamazaki2011}.
In the paper \cite{Nakamura2014}, Nakamura discussed how to construct the gauge invariant
perturbation variable in the arbitrarily high order and in the arbitrary background spacetime.
The answer to the problem presented by Nakamura will be given by our present paper. 
By using the infinitesimal gauge transformation and the application of induction 
using the concept of "the background like object", we succeed in giving the definition and the proof
of the gauge invariant perturbation variable in the most general situation.
In the paper \cite{Hamazaki2011}, the physical laws such as the Einstein evolution equations,
the Einstein constraint equations and the metric junction conditions are shown 
to be described in the manifestly gauge invariant form, more concretely speaking, to
be described in the form in which the polynomial written in terms of the single time
background/scalar like objects is vanishing.
The gauge transformation of the physical quantity is written in terms of the Lie derivatives
of the physical quantities.
The scalar like object is defined as the physical quantity whose Lie derivative is the 
same as that of the scalar quantity in the zeroth order of the gradient expansion.
The background like object is defined as the physical quantity whose gauge transformation
is the same as that of the background quantity, that is, whose gauge transformation do not
depend on the perturbation part of the gauge transformation in the zeroth order of the 
gradient expansion \cite{Hamazaki2011}.

Recently the many step reheating which contains an arbitrary large number of energy transfer
processes \cite{Elliston2014} \cite{Meyers2014} \cite{Leung2012} has been discussed.
The solution of the many step reheating based on the junction model is written by
the single time background/scalar like objects at the initial time $t_0$,  at the many times 
specifying the time of the energy transfer from the $3$ fluid representing the 
oscillation energy of each scalar field into the $4$ fluid 
$t_i$, ($i=1,2,\cdots, M$), and at the final time $t$.
(In the present paper, we call the fluid diluted in proportion of the reciprocal of $a^k$ where 
$a$ is the scale factor the $k$ fluid from now on. 
For example, the $3$ fluid, the $4$ fluid are the dust, the radiation, respectively.)
The purpose of this paper is to write the solutions of the physical laws depending upon many times
in the manifestly gauge invariant manner.
We can notice that even the solution of the simplest case contains at least two times such as the initial
time and the final present time, as demonstrated using the toy model in the appendix \ref{apdxconcrete}.
The solution expresses the present value of the physical quantity in terms of the past values of 
the physical quantities.
Therefore it is desirable that the manifestly gauge invariant perturbation of the solution 
should express the gauge invariant perturbation variables at the present time in terms of the 
gauge invariant perturbation variables at the past times.
For this purpose, we need to set time slices, the space coordinate threading except in the zeroth order
of the gradient expansion, as for the present time, all the past times appearing in the solution. 
We seek for defining the derivative operator operating on the background solution, generating  
the gauge invariant perturbation variables related with the time slices explained above
from the background physical quantities at all the times appearing in such background solution.
Our answer to this problem is the many time background/scalar like object.
In the present paper, the definition of the background/scalar like object is extended from the
single time case in the previous papers \cite{Hamazaki2008.2} \cite{Hamazaki2011} into the many 
time case by which we can treat the many step reheating as an extreme case.
By the concept of many time background/scalar like object,
the solution of the physical law becomes manifestly gauge invariant.

The physical quantities that we will seek are the Bardeen parameters $D(\rho)^k \ln{a}$
representing the $k$-th order perturbation of the logarithm of the scale factor $a$ in the 
time slice where the energy density $\rho$ is not perturbed.
$k$ is the natural number and the operator $D$ will be explained in the next section.
In order to know the Bardeen parameters from this definition, the dynamics needs to be solved
with $\rho$ as the evolution parameter explicitly.
But this cannot be realized except in the few trivial cases.
On the other hand, in the explicit way, we can easily solve the multiple slow rolling scalar fields system
using  the $\tau$ function introduced in the paper \cite{Hamazaki2008.2} as the evolution parameter,
and we can easily solve the multiple fluids system by using
the scale factor $a$ or the energy density of the either component fluid $\rho_l$ ($l=3$ or $4$)
as the evolution parameter, respectively.
Then we need to change the base representing the time slice from $\rho$ to another base such as
$\tau$, $a$, $\rho_l$ and so on.
This process will be possible by the formula of the change of the base introduced in the section
\ref{mainchangebase}.
The application of the change of the base will be discussed in the section \ref{mainchangebase} for the 
slow rolling multiple scalar fields system, in the appendix \ref{apdxchangebase} for the multiple 
fluids system, respectively

The reason why the many step reheating is considered can be explained in the following way.
We consider the inflation caused by the multiple slow rolling scalar fields.
The contribution to the Bardeen parameter at the end of the inflation is caused 
only from the adiabatic component of the multiple slow rolling scalar fields as a whole.
If after the inflation each component behaves in the same way,
all the information of the energy composition, that is the energy ratios, of the multiple 
scalar fields is lost forever.
We consider the possibility in which  each fluid component originating
from each scalar field component behaves in the different way, that is, for each component 
the time at which the transfer from the $3$ fluid to the $4$ fluid occurs is different.
In such case, we clarified which of entropic information such as the energy ratios of the multiple 
scalar fields, has influence on the present time Bardeen parameter.

This paper is organized as follows.
In section $2$, we give the definitions and the proofs of the gauge invariance 
of the gauge invariant perturbation variable in the arbitrarily high order and 
in the arbitrary background universe, and of the many time background/scalar like object 
in the zeroth order of the gradient expansion with the spatially flat homogeneous isotropic universe
being the background spacetime.
The solutions of the physical laws are interpreted in terms of the many time background/scalar 
like objects.
The background/scalar like object has the scalar like object as the base,
and the background/scalar like object as the true value and have the meaning as the 
fluctuation of the true value in the time slice in which the base is constant
\cite{Hamazaki2008.2} \cite{Hamazaki2011}.
The definitions of the base and the true value are given in the section $2$.
We discuss the relation between the definition of our gauge invariant perturbation variable 
and the definition of the gauge invariant perturbation variable which we can obtain by
generalizing  that given in the paper by Malik, Wands \cite{Malik2009} into the arbitrary order.

In section $3$, the formula on transform of the base of the background/scalar like objects
and its application is given.
The sections $4$, $5$ treats the many step reheating as a successful application of our many time
background/scalar like object.
In section $4$, we discuss the legitimacy of the junction model of the many step reheating.
In section $5$, we give new parametrization of the many step reheating and discuss its
remarkable property, the property of oblivion which brings about the large simplification
in calculating the Bardeen parameters in the many step reheating.  
By the property of oblivion, we mean that the universe forgets all the perturbations generated
by all the fluctuation processes before the prominent fluctuation generation process the nearest
to the present time.
Strictly speaking, in the Bardeen parameter in the many step reheating system, 
only the fluctuation of the $3$ fluid governing the cosmic energy lastly and 
the fluctuation of the decay constant making this $3$ fluid decay into the $4$ fluid are left.
The section $6$ is devoted to the summary and the discussion.

\section{interpretation of the solutions in terms of\\ 
the many time background/scalar like objects}

In order to fix the terminology, we shortly review the previous papers 
\cite{Hamazaki2008.2} \cite{Hamazaki2011}.
We consider the universe perturbed by the perturbation parameter $\lambda$: $M_{\lambda}$ and 
assume that the physical universe is $M_{\lambda =1}$.
For the physical quantity $A$ also, we consider the physical quantity perturbed by the perturbation 
parameter $\lambda$: $A(\lambda)$.
The $k$-th  differential coefficient of the physical quantity $A$ at $\lambda=0$ corresponds
to the $k$-th order perturbation of $A$:
\begin{equation}
 \frac{d^k}{d \lambda^k} A \Bigg|_{\lambda=0} \leftrightarrow \delta^k A.
\end{equation}
Since we want to consider that the physical quantity $A(\lambda)$ is the function of the point of the 
perturbed manifold $M_{\lambda}$, we coordinatize the perturbed universe $M_{\lambda}$ by the
coordinate of the unperturbed universe $M_{\lambda=0}$. 
The rule of the point identification between the perturbed universe $M_{\lambda}$ and the unperturbed 
universe $M_{\lambda=0}$ is called the gauge choice.
How to give the coordinates of the unperturbed universe $M_{\lambda=0}$ 
to all the points of the perturbed universe $M_{\lambda}$ is arbitrary and change of the the point 
identification between the perturbed universe $M_{\lambda}$ and the unperturbed 
universe $M_{\lambda=0}$ is called the gauge transformation, strictly speaking the gauge transformation 
of the second kind. 
The gauge transformation of an arbitrary physical quantity $A$ generated by 
$T= T^{\mu}(t,\bm{x}) \partial_{\mu}$ where $\bm{x}$ will be suppressed from now on,
is given by
\begin{equation}
 \frac{d}{d \mu} A = L(T) A,
\label{gaugetrans} 
\end{equation}
where $\mu$ is the gauge parameter by which $A(\mu=0)$ represents the original quantity
and $A(\mu=1)$ represents the gauge transformed quantity, and $L(T)$ is the Lie derivative
with respect to $T= T^{\mu} \partial_{\mu}$.
Solving the differential equation  (\ref{gaugetrans}) gives
\begin{equation}
 A(\mu=1) = \exp{[\mu L(T)]} A(\mu=0)
\end{equation}
We assume that $T= T^{\mu} \partial_{\mu}$ generating the Lie derivative, that is
the gauge transformation, depends upon the supposed perturbation
parameter $\lambda$ as well as the physical quantity $A$.
Since the gauge transformation is the change of the point identification between the perturbed universe 
$M_{\lambda}$ and the unperturbed universe $M_{\lambda=0}$,  it is self-evident that
$T(\lambda = 0) =0$.
By performing the $k$ times differentiations on the both hand sides of (\ref{gaugetrans}),
we can get the gauge transformation of the $k$-th order perturbation of $A$.
$A|_{\lambda = 0}$ corresponds to the background part of $A$ and is evidently gauge invariant
since $T(\lambda = 0) =0$. 
Since the gauge transformation of the $k$-th order perturbation of $A$,
$d^k A / d \lambda^k |_{\lambda=0}$ depends upon the Lie derivatives with respect to the 
$l$-th order perturbation of $T$, $d^l T / d \lambda^l |_{\lambda=0}$
where $l \le k$, the $k$-th order perturbation of $A$ is not gauge invariant, in general.
But by taking the appropriate combination of perturbation variables $B$
although it is constructed from the perturbation variables,
we can make the gauge transformation of $B$ not depend upon $d^l T / d \lambda^l |_{\lambda=0}$ 
so that
\begin{equation}
 \frac{d}{d \mu} B = L(T) B.
\end{equation}
The combination of the perturbation variables $B$
whose gauge transformation is the same as that of the background quantity, that is 
which is gauge invariant, although it is constructed from the perturbation variables,
is called the background like object.

We consider as the background universe the spatially flat homogeneous, isotropic universe,
that is the spatially flat Friedmann Robertson Walker universe.
The Lie derivative used in the definition of the gauge transformation has a different 
expressions depending upon the type of the transformation property against the general 
coordinate transformation, that is scalar, vector, tensor, of the physical quantity 
operated on.
The physical quantity $S$ which has the Lie derivative which is the same as that of 
the scalar except terms which can be neglected in the zeroth order of 
the gradient expansion, is called the scalar like object:
\begin{equation}
 L(T) S = T^0 \frac{\partial S}{\partial t} + O(\epsilon^2),
\end{equation}
where $\epsilon$ is the infinitesimal quantity of order of the spatial derivative,
that is the wavenumber.
In the zeroth order of the gradient expansion, it is confirmed in the paper 
\cite{Hamazaki2011} that all the physical laws such as the Einstein evolution equations, 
the Einstein constraint equations, the metric junction conditions can be expressed by 
the scalar like objects only.

We mean the quantity which is the background like object and the scalar like object 
by the background/scalar like object.
For the scalar like object $S$, and the background/scalar like object $X$,
we define the derivative operator $D(S)$ by
\begin{equation}
 D(S) X = \left[ \frac{d}{d \lambda} - 
  \left( \frac{d S}{d \lambda} \Big/ \frac{d S}{d t} \right) \frac{d}{d t} \right] X.
\label{singletimebs}
\end{equation}
Under the above definitions, we can derive the following theorems \cite{Hamazaki2011}. 
\paragraph{Theorem}
For the scalar like object $S$, and the background/scalar like object $X$,
$D(S) X$ is the background/scalar like object.

\vspace{0.5\baselineskip}
By using the above theorem inductively, we obtain
\paragraph{Theorem}
For the scalar like objects $S_1$, $S_2$, and for an arbitrary natural number $k$,
$D(S_1)^k S_2$ is the background/scalar like object.

\vspace{0.5\baselineskip}
In $D(S_1)^k S_2$,  $S_1$, $S_2$ are called the base, the true value, respectively,
borrowing the terminology of the logarithm.
$D(S_1)^k S_2$ is the background/scalar like object corresponding to the $k$-th
order perturbation of $S_2$ in the time slice where $S_1$ is constant.

As for the perturbation of the physical laws, we can confirm
\paragraph{Theorem}
The perturbation of the physical laws can be expressed in the form that the polynomial 
of the background/scalar like objects vanishes.
Therefore the perturbation of the physical laws is gauge invariant.
\vspace{0.5\baselineskip}

Next we give the construction of the gauge invariant variable for the arbitrarily high orders 
and in the arbitrary background spacetime.
The part from this to the end of Theorem $G2'$ and the end part in this section $2$ are isolated and 
discuss the gauge invariant perturbation theory with the "general" spacetime being the background spacetime 
for the arbitrarily high orders differently from the other part treating the long wavelength, 
that is the zeroth order of the gradient expansion.
The discussion of this part applies to all the theories with the general covariance including the general relativity
as the special case.
Then we would like the reader to read the Lie derivative $L(T)$ generated by the vector $T$ in this part  
as the Lie derivative in the "general" spacetime.
After the end of Theorem $G2'$, the concept of the base which will be introduced in Definition  $G4$ 
in the general spacetime will usefully be applied to the zeroth order of the gradient expansion and will be 
generalized into the many time case 
in order to construct the manifestly gauge invariant expression of the solution.
In the pioneering  paper \cite{Nakamura2014}, Dr. Nakamura tried to give the definition of the gauge invariant
perturbation variable for the arbitrarily high orders.
In the appendix \ref{apdxnakamura}, his discussion will be explained with the comparison with my present work.
Since K.Nakamura considers the gauge transformation separated by the finite gauge parameter difference, 
that is $\mu=0$ and $\mu =1$, his calculation becomes rather complicated and too difficult to be performed 
for the arbitrarily high orders.
For the arbitrarily high orders, he was obliged to introduce some conjectures which cannot be proved, 
in particular Conjecture $4.1$ in the paper \cite{Nakamura2014}.
Against his work, we will consider the infinitesimally separated gauge transformation.
That is we consider the derivative with the gauge parameter $d / d \mu$.
By this idea, the construction of the gauge invariant perturbation variable in the general spacetime
for the arbitrarily high orders will be made possible. 
Following is our answer to the question presented by Dr. Nakamura in the paper \cite{Nakamura2014}.

First we give some definitions.

\paragraph{Definition $G1$}
The physical quantity $X$ which has the following $\mu$ derivative as
\begin{equation}
 \frac{d X}{d \mu} = L(T) X,
\end{equation}
is called the background like object.
\vspace{0.5\baselineskip} 

$G$ in Definition $G1$ is abbreviation of  "general" which means the general background spacetime.
$L(T)$ implies the Lie derivative generated by the vector field $T$.

\paragraph{Definition $G2$}
The vector quantity $V^{\mu}$ which has the following $\mu$ derivative as
\begin{equation}
 \frac{d V^{\mu}}{d \mu} = \frac{d T^{\mu}}{d \lambda}+ L(T) V^{\mu},
\end{equation}
is called the gauge vector.
Please often notice that $L(T) V^{\mu} = [T, V]^{\mu}$ where $[A, B]$ is the commutator.
\vspace{0.5\baselineskip}

\paragraph{Definition $G3$}
When the differential equation
\begin{equation}
 L(V) S = U,
\end{equation}
is solved uniquely in terms of $V^{\mu}$, the set of the physical quantities $S$ is called
invertible.
\vspace{0.5\baselineskip}

\paragraph{Theorem $G1$}
When $X_n$, $V_n$ are the background like object, the gauge vector, respectively,
$X_{n+1}$ defined by
\begin{equation}
 X_{n+1} = \Biggl[\frac{d}{d \lambda} -L(V_n) \Biggr] X_n,
\end{equation}
is the background like object.
\vspace{0.5\baselineskip}

\paragraph{Proof of Theorem $G1$}
The simple calculation gives
\begin{equation}
 \frac{d}{d \mu} X_{n+1} = L(T) X_{n+1} +
 L\Biggl(  \frac{d V^{\mu}}{d \mu} - \frac{d T^{\mu}}{d \lambda}- L(T) V^{\mu}  \Biggr)
 Y_n.
\end{equation}
Therefore $X_{n+1}$ is the background like object.

\paragraph{Corollary $G1$}
When $X$, $V$ are the background like object, the gauge vector, respectively,
$X_n$ defined by
\begin{equation}
 X_n = \Biggl[\frac{d}{d \lambda} -L(V) \Biggr]^n X,
\end{equation}
is the background like object.
\vspace{0.5\baselineskip}

\paragraph{Theorem $G2$}
When the set of the background like objects $S_n$ is invertible,
$V_n$ defined as the solution of the differential equation 
\begin{equation}
 \Biggl[\frac{d}{d \lambda} -L(V_n) \Biggr] S_n =0,
\label{definitionofbase}
\end{equation}
is the gauge vector.
\vspace{0.5\baselineskip}

\paragraph{Proof of Theorem $G2$}

By differentiating  the equation (\ref{definitionofbase}) with respect to $\mu$,
we obtain 
\begin{equation}
 L(Q_n) S_n =0,
\label{gaugevectorcondition}
\end{equation}
where 
\begin{equation}
 Q_n^{\mu} : =  \frac{d V_n^{\mu}}{d \mu} - \frac{d T^{\mu}}{d \lambda}- L(T) V_n^{\mu}.  
\label{definitionofQ}   
\end{equation}
Since the set of the background like objects $S_n$ is invertible, we can conclude $Q_n =0$
which means that $V_n^{\mu}$ is gauge vector.

\paragraph{Definition $G4$}
$V_n$ defined as the solution of the differential equation (\ref{definitionofbase})
is called the gauge vector with the base $S_n$.
\begin{equation}
 D(S_n): = \frac{d}{d \lambda} -L(V_n),
\end{equation}
is called the covariant derivative with the base $S_n$.
\vspace{0.5\baselineskip}

In the general spacetime, we cannot often write the analytic expression to the equation  (\ref{definitionofbase})
explicitly.
But in the practical case, it is sufficient to solve the equation (\ref{definitionofbase}) perturbatively, that is 
to write the solution as the power expansion with respect to the perturbation parameter $\lambda$, around the point
where the perturbation parameter $\lambda$ is vanishing.
In this case, the theorem $G2$ is changed into the following form.
It can be proved inductivlely  by differentiating the equation (\ref{gaugevectorcondition}) $k$ times 
with respect to $\lambda$ and putting $\lambda = 0$ afterward where $k$ is an nonnegative integer.     

\paragraph{Theorem $G2'$}
When the set of the background like objects at $\lambda=0$, that is $S_n |_{\lambda = 0}$ is invertible,
$V_n$ defined as the solution of the equation (\ref{definitionofbase}) satisfies 
$d^{k} / d \lambda^k \cdot Q_n |_{\lambda =0} =0$ for an arbitrary nonnegative integer $k$ where $Q_n$ is defined 
by the equation (\ref{definitionofQ}).
That is until the sufficiently high order with respect to the perturbation parameter $\lambda$, $Q_n$ is kept vanishing.
\vspace{0.5\baselineskip}

In the appendix \ref{apdxgeneralbase}, we discuss the concept of the base in the more general setup using the
background homogeneous isotropic universe without using the gradient expansion unlike the main body of this paper. 

From now on, we specialize the discussion into the zeroth order of the gradient expansion around the background
spatially flat homogeneous isotropic universe.
In case of the single time background/scalar like object (\ref{singletimebs}), the gauge vector is given by 
\begin{equation}
V^0 = \frac{d S}{d \lambda} \Big/ \frac{d S}{d t}. 
\end{equation}

In this paper, in order to treat the solutions of the physical laws in the manifestly 
gauge invariant manner, we generalize the background like object and the scalar like
object into the many time case.
The expression of the solution of the cosmological situations contains many times, for 
example, the initial time, the final present time, the times setting the metric junctions
such as the energy transfer from the dust to the radiation appearing in the reheating. 
The situation where we have to treat the many time case is explained easily in the appendix 
\ref{apdxconcrete} where the many time case is explained using the concrete toy model.  
Since for different times, how to set the time slices can be arbitrary and different from 
each other, this discussion is sufficiently meaningful.
The meaning of the many time background like object is self-evident.
The many time scalar like object $S(t_1, \cdots, t_n)$ is defined to be the quantity
which has the Lie derivative 
\begin{equation}
 L(T) S = \sum_i T^0(t_i) \frac{\partial S}{\partial t_i},
\end{equation}
in the zeroth order of the gradient expansion.
From now on, $\partial$ is often simply expressed as $d$ for notational simplicity
when it can induce no confusion.
A polynomial constructed from single time scalar like objects at plural different  
times is self-evidently a many time scalar like object.

We consider the set of $n$ many time scalar like objects $Z_i$, $\{ Z_i \}^n_{i=1}$ 
which satisfies the regularity condition $\det{\partial Z_j / \partial t_i} \neq 0$,
that is, the condition that the base $\{ Z_i \}^n_{i=1}$ is invertible,
which will not be mentioned from now on.
We describe as $d t_i / d \lambda$ the solution of the equation system defined by
\begin{equation}
 \frac{d Z_j}{d \lambda} + \sum_{i} \frac{d t_i}{d \lambda} \frac{d Z_j}{d t_i} =0, 
\end{equation}
where $d t_i / d \lambda$ are simply symbols which describe the solution of the above 
equation system and they does not necessarily imply that $t_i$ depend upon $\lambda$.
We consider the set of the time slices $Z_i =0$.
$d t_i / d \lambda$ are given by
\begin{equation}
 \frac{d t_i}{d \lambda} = - \sum_j \frac{d Z_j}{d \lambda} \frac{d t_i}{d Z_j}. 
\end{equation}
$d t_i / d Z_j$ is defined as the inverse matrix of $d Z_j / d t_k$.
For the set of many time scalar like objects $\{ Z_i \}^n_{i=1}$ and a many time 
background/scalar like object $X$, we define the derivative operator 
$D(Z_1, \cdots, Z_n)$ by
\begin{equation}
 D(Z_1, \cdots, Z_n) X = \left( \frac{d}{d \lambda} + 
  \sum_i \frac{d t_i}{d \lambda} \frac{d}{d t_i} \right) X 
  = \left( \frac{d}{d \lambda} - 
  \sum_{ij} \frac{d Z_j}{d \lambda} \frac{d t_i}{d Z_j} \frac{d}{d t_i} \right) X.   
\end{equation}
Under the above definitions, we obtain the following theorems:

\paragraph{Theorem $2.1$}
For the set of many time scalar like objects $\{ Z_i \}^n_{i=1}$, the many time 
background/scalar like object $X$, $D(Z_1, \cdots, Z_n) X$ is a many time background/scalar 
like object.

\vspace{0.5\baselineskip}
For the proof, please see appendix \ref{apdxproof2.1}.
In the many time background/scalar like object $D(Z_1, \cdots, Z_n) X$, we can see that the gauge vector is given by
\begin{equation}
V_{i} = \sum_{j} \frac{d Z_j}{d \lambda} \frac{d t_i}{d Z_j}. 
\end{equation}
By using the above theorem inductively, we obtain

\paragraph{Theorem $2.2$}
For the set of many time scalar like objects $\{ Z_i \}^n_{i=1}$, the many time 
scalar like object $X$ and an arbitrary natural number $k$, $D(Z_1, \cdots, Z_n)^k X$ is 
a many time background/scalar like object.

\vspace{0.5\baselineskip}
The set of many time scalar like objects $\{ Z_i \}^n_{i=1}$, the many time scalar like
object $X$ in $D(Z_1, \cdots, Z_n)^k X$ will be called the base, the true value, respectively, 
in the same manner as in the case of the single time background/scalar like object. 
In other words known better,
$D(Z_1, \cdots, Z_n)^k X$ is the gauge invariant variable representing the $k$-th order 
perturbation of $X$ under the set of the time slices where $Z_1, \cdots, Z_n$ are constant.

We consider the solution of the physical laws.
The solution can be interpreted as the relation between the scalar like objects at the 
initial time $t_0$ and the scalar like objects at the final time $t$.
Therefore the solution can be written in the form where many time scalar like objects at 
arbitrary times $t_0$, $t$, $X_A (t_0, t)$, where in order to describe the solution plural
(distinguished by indices $A$) independent many time scalar like objects are used, vanish:
\begin{equation}
 X_A (t_0, t) =0.
\end{equation}
By operating $D(Z_0, Z)^k$ where $k$ is an arbitrary natural number 
on the above solution, for arbitrary independent many time scalar like 
objects $Z_0$, $Z$ both of which depend upon $t_0$, $t$, we can obtain the manifestly 
gauge invariant $k$-th order perturbation expression of this solution.

In order for the reader to understand the situation where the physical solution is put,
in appendix \ref{apdxconcrete}, we give the explanation easier to understand as for the perturbation solution 
in terms of the many time background/scalar like object using the simple concrete toy model.

As a simple method for constructing the solution from the inflationary stage to the 
present time based on the inflationary theory \cite{Sato1981} \cite{Guth.1981}, 
the junction model exists.
In this junction model, changes which happen in a comparatively short time such as
the changes from the slow rolling phase 
in which phase scalar fields can be treated by the $\tau$ function \cite{Hamazaki2008.2}
\cite{Hamazaki2011} into the oscillatory phase in which phase scalar fields can be described 
by the $3$ fluids in a sufficient accuracy \cite{Kodama1996} \cite{Hamazaki2002} \cite{Hamazaki2004},
the energy transfers from the $3$ fluid into the $4$ fluid in the reheatings, are treated 
as metric junctions at a single time \cite{Israel1966} \cite{Deruelle1995} \cite{Martin1998}
\cite{Copeland2007} \cite{Mukohyama2000} \cite{Hamazaki2011}.
The spacelike surface on which each metric junction is performed is defined 
by 
\begin{equation}
 C_i =0, \label{metricjunctionscalarlikeobj}
\end{equation}
where $C_i$ is a single time scalar like object at a single time $t=t_i$ \cite{Hamazaki2011}.
For example, as for a change from the slow rolling phase into the oscillatory phase of the scalar 
field, the surface  on which the metric junction is performed is defined by
\begin{equation}
 \left( \frac{1}{\alpha} \frac{\dot{a}}{a} \right)^2 -m^2 =0,
\end{equation}
where $\alpha$ is the lapse, $a$ is the scale factor, the dot on $a$ implies the time $t$ derivative,
and $m$ is the mass of the scalar field.
As for the energy transfer from the $3$ fluid into the $4$ fluid, the surface on which
the metric junction is performed is defined by
\begin{equation}
 \left( \frac{1}{\alpha} \frac{\dot{a}}{a} \right)^2 -\Gamma^2 =0,
\end{equation}
where $\Gamma$ is the decay constant of the $3$ fluid to the $4$ fluid, 
which can be a function of another fluctuating scalar field in the modulated reheating 
scenario \cite{Sasaki1991} \cite{Dvali2004}.
In such case as contains many metric junctions, the expression of the solution can be 
written by plural many time scalar like objects $X_A$ as
\begin{equation}
 X_A (t_0, t_1, \cdots, t_M, t) =0,\label{solutionmetricjunctions}
\end{equation}
where $X_A$ depends on not only the initial time $t_0$, the present time $t$, 
but also the times $t_i$, ($i=1, \cdots, t_M$) on which the metric junction is performed.
In this expression, while the initial time $t_0$ and the present time $t$ are arbitrary,
$t_i$ are not arbitrary and depend upon the supposed perturbation parameter $\lambda$.
By Eq.(\ref{metricjunctionscalarlikeobj}), $d t_i / d \lambda$ is calculated by
\begin{equation}
 \frac{d C_i}{d \lambda} + \frac{d t_i}{d \lambda} \frac{d C_i}{d t_i} =0.
\end{equation}
Therefore operating $D(Z_0, Z)^k$ on Eq.(\ref{solutionmetricjunctions}) gives
\begin{equation}
 D(Z_0, C_1, \cdots, C_M, Z)^k X_A =0,
\end{equation}
as the $k$-th order perturbation part of the solution.
Notice that while $Z_0$, $Z$ are arbitrary, $\{C_i \}^M_{i=1}$ cannot be arbitrary, 
that is must be the set of scalar like objects defining the surfaces on which the metric 
junctions are performed.
This fact will be used in the manifestly gauge invariant treatment of the many step 
reheating discussed later.

From the present point to the end of this section, we consider the arbitrary spacetime
as the background spacetime.
In this section, we gave the way to construct the gauge invariant variable in an arbitrary high order and 
in the general background spacetime.
On the other hand, in the paper \cite{Malik2009},
Malik, Wands proposed another way to construct the gauge invariant variable in the second
order cosmological perturbation theory, in which they use the gauge transformed quantity $\exp{[- L(\tilde{V})]} X$
where $\tilde{V}$ is the vector corresponding with the gauge vector of the second type defined by us a little later 
and $X$ is an arbitrary background like object.
Next we will extend the way proposed by Malik, Wands into an arbitrary order and an arbitrary background 
spacetime, and will call this extension the MW like construction.
The MW construction becomes transparent by watching 
the generating function generating all order perturbations $\exp{[- L(\tilde{V})]} X$.

\paragraph{Definition $GMW.1$}
The vector quantity $\tilde{V}^{\mu}$ which has the following $\mu$ derivative as
\begin{equation}
 \sum^{\infty}_{k=0} \frac{1}{(k+1)!} 
\underbrace{[ \tilde{V} [ \tilde{V} [ \cdots [ \tilde{V} }_{k} ,- \frac{d \tilde{V} }{d \mu}] \cdots ]]]
+ T=0,
\end{equation}
is called the gauge vector of the second type.
\vspace{0.5\baselineskip}

\paragraph{Theorem $GMW.1$}
When the set of the background like objects $S$ is invertible,
$\tilde{V}$ defined as the solution of the differential equation 
\begin{equation}
 \exp{[- L(\tilde{V})]} S =S|_{\lambda =0},
\label{defvectorsecondtype}
\end{equation}
is the gauge vector of the second type.
\vspace{0.5\baselineskip}

\paragraph{Proof of Theorem $GMW.1$}
By differentiating Eq.(\ref{defvectorsecondtype}) with respect to $\mu$ with using the equality
\begin{equation}
 \exp{[ L(\tilde{V})]}  \frac{d}{d \mu}  \exp{[- L(\tilde{V})]}     + L(T) = L(\tilde{Q}),
\end{equation}
where
\begin{equation}
 \tilde{Q} :=  \sum^{\infty}_{k=0} \frac{1}{(k+1)!} 
\underbrace{[ \tilde{V} [ \tilde{V} [ \cdots [ \tilde{V}}_{k} ,- \frac{d \tilde{V} }{d \mu}] \cdots ]]] + T,       
\end{equation}
we obtain

\begin{equation}
 \exp{[- L(\tilde{V})]} L(\tilde{Q}) S =0.
\end{equation}
Since the set of the background like objects $S$ is invertible, we obtain $\tilde{Q} = 0$ implying
$\tilde{V}$ is the gauge vector of the second type.

\paragraph{Theorem $GMW.2$}
When $X$ is the background like object, and $\tilde{V}$ is the gauge vector of the second type,
$\exp{[- L(\tilde{V})] X}$ is gauge invariant. 
\vspace{0.5\baselineskip}

\paragraph{Proof of Theorem $GMW.2$}
Since $\tilde{Q} = 0$,
\begin{equation}
 \frac{d}{d \mu} \{ \exp{[- L(\tilde{V})]} X \} =  \exp{[- L(\tilde{V})]} L(\tilde{Q}) X  =0.
\end{equation}
\vspace{0.5\baselineskip}

The gauge invariant perturbation variable given as the generalization to the arbitrarily high order of
the gauge invariant perturbation variable defined by Malik, Wands in the second perturbation theory 
corresponds with
\begin{equation}
 \frac{d^k}{d \lambda^k} \{ \exp{[- L(\tilde{V})]} X \} \Bigg|_{\lambda =0 } \quad (k=1,2,\cdots).
\end{equation}

We would like to establish the equivalence between our construction presented in the preceding part
of this section and the MW like construction.  
For this purpose, the following theorem is necessary. 

\paragraph{Theorem $GMW.3$}
$V$ is the vector field whose $(k-1)$-th coefficient of the Taylor expansion is $V_k$ ($k=1,2,\cdots$) :
\begin{equation}
 \frac{d^{k-1}}{d \lambda^{k-1}} V \Bigg|_{\lambda =0} = V_k \quad (k = 1,2, \cdots).
\end{equation}
$\tilde{V}$ is the vector field which can be expanded as 
\begin{equation}
 \tilde{V} =  \sum^{\infty}_{k=1} \frac{\lambda^k}{k!}  \tilde{V}_k.
\end{equation}
$X$ is the background like object which can be expanded as
\begin{equation}
 X =   \sum^{\infty}_{k=0}  \frac{\lambda^k}{k!} \frac{d^{k}}{d \lambda^{k}} X \Bigg|_{\lambda =0}
\end{equation}
For an arbitrary vector field $V$, there exists an vector field $\tilde{V}$ which satisfies
\begin{equation}
 \sum^{\infty}_{k=0} \frac{\lambda^k}{k!} \left[\frac{d}{d \lambda} -L(V) \right]^k X \Bigg|_{\lambda =0} =
 \exp{[- L(\tilde{V})]} X,
\label{equivalencegenerator}
\end{equation}
and satisfies
\begin{equation}
 \tilde{V}_k = V_k + R_k \quad(k=1,2,\cdots),
\label{equivalencerelation}
\end{equation}
where $R_k$ is the sum of the terms in which all the products are written by the commutators 
with respect to $V_l$ ($l \le k-1$),  and which gives the one to one correspondence independent of 
the arbitrary background like object $X$ between  vector fields $\tilde{V}$ and $V$.
\vspace{0.5\baselineskip}

In fact, for small natural numbers $k$, direct calculations gives
\begin{align}
 \tilde{V}_1 &= V_1,\\
 \tilde{V}_2 &= V_2,\\
 \tilde{V}_3 &= V_3 - \frac{1}{2} [V_1, V_2],\\
 \tilde{V}_4 &= V_4 - [V_1, V_3],\\
 \tilde{V}_5 &= V_5 - \frac{3}{2} [V_1, V_4]  - [V_2, V_3]  + \frac{1}{2} [[V_1, V_2], V_2] 
+ \frac{1}{6} [V_1, [V_1, V_3]]  + \frac{1}{6} [V_1, [V_1, [V_1, V_2]]].
\end{align}
Since seen from the proof in appendix \ref{apdxproofG6.3} we could prove that the above theorem is right by 
the Baker Campbell Hausdorff formula, 
we can show that our construction in this section and the MW like construction are equivalent in the 
generating function level, that is, in an arbitrary finite order perturbation.
In our definition of the gauge invariant perturbation variable in the preceding part of this section, 
the gauge vector $V$ is given by the solution $V$ to the equation
\begin{equation}
\left[ \frac{d}{d \lambda} - L(V) \right] S =0,
\end{equation}
for some background like object $S$, and for an arbitrary real value $\lambda$.
Therefore $V$ satisfies   
\begin{equation}
 \left[ \frac{d}{d \lambda} - L(V) \right]^k S =0,
\end{equation}
for all the natural numbers $k$.
Then our gauge invariant variable is given by
\begin{equation}
  \left[ \frac{d}{d \lambda} - L(V) \right]^l X \Bigg|_{\lambda=0} 
\end{equation}
whose $V$ satisfies
\begin{equation}
 \left[ \frac{d}{d \lambda} - L(V) \right]^k S \Bigg|_{\lambda =0} =0,
\end{equation}
where $l$, $k$ are arbitrary natural numbers.
On the other hand, the MW gauge invariant variable is given by the coefficient of $\lambda^l$ of
\begin{equation}
 \exp{[- L(\tilde{V})]} Y
\end{equation}
under the gauge vector of second type satisfying
\begin{equation}
 \exp{[- L(\tilde{V})]} S =S|_{\lambda = 0}.
\end{equation}
By the one to one correspondence between $\tilde{V}$ and $V$ given by the above proposition, 
we can see that our definition and the MW definition correspond completely.
The difference between the both is only the difference of the parametrization of the gauge vector.

\section{formula on transform of the base of \\ the background/scalar like objects}
\label{mainchangebase}

We prove two useful formulas on the derivative operator $D(Z_1, \cdots, Z_n)$
which generates from an old many time background/scalar like object 
another new many time background/scalar like object.
For a set of $n$ independent many time scalar like objects at $\{t_i \}^n_{i=1}$,
$\{Z_i \}^n_{i=1}$, and a many time background/scalar like object $X$, 
we define the derivative operator $d / d Z_i$ by
\begin{equation}
 \frac{d}{d Z_i} X = \sum_{k} \frac{d t_k}{d Z_i} \frac{d X}{d t_k},
\end{equation}
which is also the derivative operator generating from an old many time 
background/scalar like object another new many time background/scalar like object 
for the reason analogous to the latter half of Appendix \ref{apdxproof2.1}.

The above two derivative operators satisfy the following commutation relation:
\paragraph{Theorem $3.1$}
For two sets of the $n$ independent many time scalar like objects $\{Y_i \}^n_{i=1}$, 
$\{Z_i \}^n_{i=1}$, the following commutation relation holds: 
\begin{equation}
 \left[D(Y_1, \cdots, Y_n), \frac{\partial}{\partial Z_i} \right]
 = - \sum_j \frac{\partial}{\partial Z_i} [D(Y_1, \cdots, Y_n) Z_j] \cdot
 \frac{\partial}{\partial Z_j}.
\end{equation}

\vspace{0.5\baselineskip}
For the proof, please see appendix \ref{apdxproof3.1}.
This formula can be used in the following way.
In using $\rho$ as the evolution parameter \cite{Hamazaki2011}, 
the left hand side of the evolution equation
is written by $d X / d \rho$ where $X$ is a single time background/scalar like object.
When we operate $D(\rho)$ on the both sides of the evolution equation in order 
to obtain the higher order of the perturbation equation, since $d / d \rho$, $D(\rho)$
commute according to the above theorem, left hand side of the perturbation evolution 
equation becomes $d D(\rho) X / d \rho$, that is, the evolution equation of the perturbation
quantity $D(\rho) X$ is obtained.

Next we can obtain the more important theorem, the formula on transform of the base of the 
many time background/scalar like object: 
 
\paragraph{Theorem $3.2$}
For two sets of the $n$ independent many time scalar like objects $\{Y_i \}^n_{i=1}$, 
$\{Z_i \}^n_{i=1}$ and a many time background/scalar like object $X$,
the following formula holds:
\begin{equation} 
 D(Y_1, \cdots, Y_n)X = D(Z_1, \cdots, Z_n)X - \sum_i D(Z_1, \cdots, Z_n)Y_i \cdot
 \frac{\partial X}{\partial Y_i} 
\end{equation}

\vspace{0.5\baselineskip}
For the proof, please see appendix \ref{apdxproof3.2}.
This formula guarantees that an arbitrary perturbation variable on the base $\{Y_i \}^n_{i=1}$
can be constructed with the perturbation variables on another arbitrary base 
$\{Z_i \}^n_{i=1}$.
In fact, as for the second order perturbation quantity, by applying the theorem $3.2$ two
times, we obtain
\begin{align} 
 D(Y_1, \cdots, Y_n)^2 X &= D(Z_1, \cdots, Z_n)^2 X - \sum_i D(Z_1, \cdots, Z_n)^2 Y_i \cdot
 \frac{\partial X}{\partial Y_i} \notag\\
&- 2 \sum_i D(Z_1, \cdots, Z_n) Y_i \cdot D(Z_1, \cdots, Z_n) \frac{\partial X}{\partial Y_i}
\notag\\
&+ \sum_{ij} D(Z_1, \cdots, Z_n) Y_i \cdot D(Z_1, \cdots, Z_n) Y_j \cdot 
 \frac{\partial^2 X}{\partial Y_i \partial Y_j}.
\end{align}
By using the theorem $3.2$ inductively, we can conclude that
for an arbitrary many time scalar like object $X$, 
and an arbitrary natural number $k$, $ D(Y_1, \cdots, Y_n)^k X$ can be written 
in the form of the polynomial constructed with several other many time scalar like objects 
with the derivative operator $ D(Z_1, \cdots, Z_n)^l$, $l \le k$ operated on.

Now we discuss the geometrical meaning of the derivative operators 
$d / d Z_i$ ($i=1, \cdots, n$) and $D(Z)$ shortly.
We consider the manifold coordinatized by the perturbation parameter $\lambda$ and 
the many times $t_i$ ($i=1, \cdots, n$).
By the definitions and theorem $3.1$, it can be verified that 
$d / d Z_i$ ($i=1, \cdots, n$), $D(Z)$ form the commutative base of the tangent vectors.
The dual base of the cotangent vectors is given by $d \lambda$, 
$d Y_i := d Y_i / d \lambda \cdot d \lambda + d Y_i / d t_j \cdot d t_j$.
From the above consideration, $D(Z)$ can be verified to be the derivative operator 
increasing $\lambda$ with fixing the values of $Z_i$ ($i=1, \cdots, n$).
This is consistent with the fact that the derivative operator $D(Z)$ is related 
with the perturbation under the set of the time slices defined by $Z_i = 0$ 
($i=1, \cdots, n$).

As the application of the formula on the transform of the base of the many time 
background/scalar like object, the system of the slow rolling scalar fields is considered.
Although the slow rolling scalar fields are governed by the evolution equation as
\begin{equation}
 \frac{d \phi_a}{d N} = - \frac{1}{\kappa^2 U} \frac{\partial U}{\partial \phi_a},
\end{equation}
where $N$ is the logarithm of the scale factor $a$, $\ln{a}$,
$\kappa$ is the gravitational constant and $U$ is the potential, in many practical cases
the solution of $\phi_a$ cannot be written as the function of $N$ manifestly.
Then following the paper \cite{Hamazaki2008.2}, in order to obtain the manifest expressions 
of $\phi_a$, $N$, by introducing the $\tau$ function $\tau$ we decompose the evolution equation as
\begin{equation}
 \frac{d \phi_a}{d \tau} = - \frac{\partial U}{\partial \phi_a}, \quad
 \frac{d N}{d \tau} = {\kappa^2 U}, 
\end{equation}
where after expressing the scalar fields $\phi_a$ as the function of $\tau$ by integrating 
the first equations, by integrating the second equation with these expressions of $\phi_a$
substituted  into the right hand side, $N$ can be written as the function of $\tau$.
Under the assumption that the potential $U$ is given by
\begin{equation}
 U = \sum_a \frac{1}{2} m^2_a \phi^2_a,
\end{equation}
where $m_a$ is the mass of the scalar field $\phi_a$, 
the scalar field $\phi_a$, the logarithm of the scale factor $N$ can be calculated as
\begin{align}
 \phi_a &= \phi_a (0) \exp{[-m^2_a (\tau - \tau(0))]},\\
 N &= N(0) + \frac{\kappa^2}{4} \sum_a \phi^2_a (0) -
 \frac{\kappa^2}{4} \sum_a \phi^2_a (0) \exp{[-2 m^2_a (\tau - \tau(0))]}.
\end{align}
Although for a natural number $k$,
the $k$-th order Bardeen parameter $\zeta_k$ defined in the paper \cite{Hamazaki2008.2}
can be obtained by applying $D(\rho, a(0))^k$ to the above expression $N$,
such obtained $\zeta_k$ contains the higher order $\tau$ fluctuations which are not related 
with the direct observations at all given by $D(\rho, a(0))^l (\tau - \tau(0))$ where 
$l$ is a natural number satisfying $l \le k$.
Please notice that $\tau$ is the mere convenient scalar like object considered in order to 
write the expressions of the solution manifestly.
In order to push the $\tau$ perturbations out of the expressions of the Bardeen parameter
$\zeta_k$, we consider the transform of the base as
\begin{equation}
 \rho, a(0) \to \tau-\tau(0), a(0).
\end{equation}
Since under the new base, for a natural number $k$, 
\begin{align}
 &D(\tau-\tau(0), a(0))^k \left( \tau-\tau(0) \right) =0, \\ 
 &D(\tau-\tau(0), a(0))^k N(0) =0, \\
 &D(\tau-\tau(0), a(0))^k \phi_a (0) = D(a(0))^k \phi_a (0),
\end{align}
$\zeta_k$ can be written with only the quantities which can be determined by the scalar fields 
dynamics in the horizon such as the scalar fields fluctuations in the flat slice.
By using the formula on the transform of the base of the many time background/scalar
like object, we obtain
\begin{equation}
 D(\rho, a(0)) N = D(\tau-\tau(0), a(0)) N -
 D(\tau-\tau(0), a(0)) \rho \cdot \frac{d N}{d \rho}, 
\end{equation}
and  
\begin{align}
 &D(\rho, a(0))^2 N = D(\tau-\tau(0), a(0))^2 N -
 D(\tau-\tau(0), a(0))^2 \rho \cdot \frac{d N}{d \rho} \notag\\
 &- 2 D(\tau-\tau(0), a(0)) \rho \cdot
D(\tau-\tau(0), a(0)) \frac{d N}{d \rho}   
+ \left[ D(\tau-\tau(0), a(0)) \rho \right]^2 \cdot \frac{d^2 N}{d \rho^2}.
\end{align}
For simplicity, we assume that the nonlinear components of the initial scalar fields
fluctuations are vanishing: $D(a(0))^k \phi_a (0) = 0$ for $k \ge 2$.
The coefficients of the first order, the second order Bardeen parameters defined by  
\begin{equation}
 D(\rho) N = N_a \cdot D(a(0)) \phi_a (0), \quad
 D(\rho)^2 N = N_{ab} \cdot D(a(0)) \phi_a (0) \cdot D(a(0)) \phi_b (0),
\end{equation}
can be calculated as
\begin{align}
 N_a &= \frac{\kappa^2}{2} \phi_a (0) -
 \frac{\kappa^2}{2} \phi_a (0) \exp{[- 2 m^2_a (\tau-\tau(0))]} \notag\\
&+
 \frac{\kappa^2}{2} m^2_a \phi_a (0) \exp{[- 2 m^2_a (\tau-\tau(0))]}
 \frac{A(2)}{A(4)},
\end{align}
and 
\begin{align}
 &N_{ab} = \frac{\kappa^2}{2} \delta_{ab} -
 \frac{\kappa^2}{2} \delta_{ab} \exp{[- 2 m^2_a (\tau-\tau(0))]}
 +
 \frac{\kappa^2}{2} m^2_a \exp{[- 2 m^2_a (\tau-\tau(0))]} \frac{A(2)}{A(4)}
 \delta_{ab} \notag\\
&+ 2 \kappa^2 \frac{1}{A(4)} 
 m^2_a \phi_a (0) \exp{[- 2 m^2_a (\tau-\tau(0))]}
 m^2_b \phi_b (0) \exp{[- 2 m^2_b (\tau-\tau(0))]} \notag\\
&- \kappa^2 \frac{A(2)}{A(4)^2} \left( m^2_a m^4_b + m^4_a m^2_b \right)
\phi_a (0) \phi_b (0)
 \exp{[- 2 m^2_a (\tau-\tau(0))]}
 \exp{[- 2 m^2_b (\tau-\tau(0))]} \notag\\
&- \kappa^2 m^2_a \phi_a (0) \exp{[- 2 m^2_a (\tau-\tau(0))]} 
           m^2_b \phi_b (0) \exp{[- 2 m^2_b (\tau-\tau(0))]}
\left[ 1 - \frac{A(2) A(6)}{A(4)^2}    \right]  
 \frac{1}{A(4)}, 
\end{align}
where
\begin{equation}
 A(2 n) := \sum_a \left( m^2_a \right)^n \phi^2_a (0) \exp{[- 2 m^2_a (\tau-\tau(0))]}.
\end{equation}
By using the coefficients $N_a$, $N_{ab}$, we can calculate the physical quantities 
which can be compared with the observations \cite{Komatsu2001} such as the non-Gaussianity
defined by
\begin{equation}
 f_{NL} = \frac{N_{ab} N_a N_b}{ \left( N_c N_c \right)^2 },
\end{equation}
the wavelength dependence of the first order Bardeen parameter defined by 
\begin{equation}
 \frac{d}{d \ln{k}} \ln{\zeta^2_1} = \frac{1}{\kappa^2 U(0)}
 \left[
-2 \frac{\sum_{ab} N_a N_{ab} U_b (0)}{\sum_a N^2_a} 
- \frac{1}{U(0)} \sum_a U_a (0)^2 
 \right], 
\end{equation}
where $U_a (0) := \partial U(0) / \partial \phi_a (0)$,
$\zeta_1$ is the first order Bardeen parameter and $k$ is the wavenumber.

\section{the junction model of the many step reheating}

As the reheating occurring after the inflation by multiple scalar fields,
we consider the many step reheating by which we mean the reheating where
the number of times of the energy transfers from the 
$3$ fluid into the $4$ fluid is arbitrarily large.
In order to characterize the properties of the plural energy transfers in the 
many step reheating more information is necessary than in the single step reheating, that is,
not only the adiabatic information such as the energy of the whole scalar fields
but also the entropic information such as the energy ratios between the scalar fields.
We treat the many step reheating by the junction model in which the energy transfers 
are treated as the metric junctions from the $3$ fluid into the $4$ fluid.
Compared with the junction model, the legitimate method which solves the system of
the fluid differential equations with decay terms will be called the DD model, with DD being
the abbreviation of the Differential equations with Decay terms.
We consider how faithfully the junction model can describe the results of the DD model,
by evaluating the difference between the solution of the DD model and that of the junction
model using $\rho$ as the evolution parameter \cite{Hamazaki2011}, more concretely speaking, 
as for only one energy transfer picked up from the many step reheating the difference of the 
both will be investigated.  
(By the way, from the reviewer, the author got to know that our junction model corresponds with
"the sudden decay approximation" in the paper\cite{Sasaki2006}.)

Although other situations can be also treated, we consider the following representative 
situation in which the solution by the DD model can be written down easily:
while the subdominant $\rho_{3r}$, $\rho_{4r}$ are the components not being changed in the 
considered time interval, $\rho_{3i}$ governing the cosmic energy is transferred into 
$\rho_{4i}$ by the $\gamma_i$ decay term.
$\rho_{3i}$, $\rho_{4i}$ obey the DDs defined by
\begin{equation}
 \frac{d \rho_{3i}}{d \rho} = \frac{1}{3 \rho_3 + 4 \rho_4} 
 \left( 3 \rho_{3i} + \frac{\sqrt{3} \gamma_i}{\kappa \rho^{1/2}} \rho_{3i} \right),
\end{equation}
\begin{equation}
  \frac{d \rho_{4i}}{d \rho} = \frac{1}{3 \rho_3 + 4 \rho_4} 
 \left( 4 \rho_{4i} - \frac{\sqrt{3} \gamma_i}{\kappa \rho^{1/2}} \rho_{3i} \right),
\end{equation}
respectively, 
where $\rho_3$, $\rho_4$ are the sums of the energy density of the $3$ fluid, the $4$ fluid 
of the total system, respectively, and $\gamma_i$ is the decay constant  
causing the energy transfer from $\rho_{3i}$ into $\rho_{4i}$.
We called "constant", although such decay constant can however 
fluctuate in the modulated reheating, but from now on without mentioning we will use the word 
"constant",.
We assume that between the initial energy density $\rho(0)$ and $\rho(1)$ defined in the 
latter half in this sentence, 
the $3$ fluid $\rho_3$, that is  $\rho_{3i}$ dominates the cosmic energy: 
\begin{equation}
 3 \rho_3 + 4 \rho_4 \cong 3 \rho,
\end{equation}
and that from $\rho(1)$ the $4$ fluid $\rho_4$, that is $\rho_{4i}$ gets to dominate 
the cosmic energy: 
\begin{equation}
 3 \rho_3 + 4 \rho_4 \cong 4 \rho,
\end{equation}
where $\rho(1)$ is defined by $\beta$, the constant of order of unity as
\begin{equation}
 \beta = \frac{\sqrt{3} \gamma_i}{2 \kappa} \frac{1}{\rho(1)^{1/2}}.
\end{equation}
Under this assumption, we can solve the above DDs analytically.
For $\rho(0) \ge \rho \ge \rho(1)$, the dominant components are given by
\begin{equation}
 \rho_{3i} = \rho_{3i} (0) \left( \frac{\rho}{\rho(0) }\right),\quad
 \rho_{4i} = 0,
\end{equation} 
and the subdominant components are given by
\begin{equation} 
 \rho_{3r} = \rho_{3r} (0) \left( \frac{\rho}{\rho(0) }\right),\quad
 \rho_{4r} = \rho_{4r} (0) \left( \frac{\rho}{\rho(0) }\right)^{4/3}.
\end{equation}
For $\rho(1) \ge \rho$, the dominant components are given by
\begin{equation}
 \rho_{3i} = \rho_{3i} (1) \left( \frac{\rho}{\rho(1) }\right)^{3/4}
 \exp{(-t + \beta)},\quad
 \rho_{4i} = \rho_{3i} (1) \left( \frac{\rho}{\rho(1) }\right) \int_{\beta}
 dt \left( \frac{t}{\beta} \right)^{1/2} \exp{(-t + \beta)},
\end{equation} 
and the subdominant components are given by
\begin{equation} 
 \rho_{3r} = \rho_{3r} (1) \left( \frac{\rho}{\rho(1) }\right)^{3/4},\quad
 \rho_{4r} = \rho_{4r} (1) \left( \frac{\rho}{\rho(1) }\right),  
\end{equation}
where $t$ is defined by
\begin{equation}
 t:= \frac{\sqrt{3} \gamma_i}{2 \kappa} \frac{1}{\rho^{1/2}}.
\end{equation}
But it is difficult to obtain the analytic expression of $N-N(0)$ by integrating 
\begin{equation}
 \frac{d N}{d \rho}= - \frac{1}{3 \rho_3 + 4 \rho_4},
 \label{integrationscalefactor}
\end{equation}
with $\rho_{3i}$, $\rho_{4i}$, $\rho_{3r}$, $\rho_{4r}$ which have been obtained in the 
above substituted to the righthand side.
Then we consider integrating the DD (\ref{integrationscalefactor}) moving to 
the junction model corresponding with the above DD model.
In the following, we construct the corresponding junction model.
First we must define the energy density specifying the metric junction surface
$\rho_{\ast}$.   
While the subdominant energy densities $\rho_{3r}$, $\rho_{4r}$ are not changed,
the counterparts of the junction model corresponding with $\rho_{3i}$, $\rho_{4i}$;
$R_{3i}$, $R_{4i}$ are changed as follows:
For $\rho(1) \ge \rho \ge \rho_{\ast}$,
\begin{equation}
 R_{3i} = \rho_{3i} (1) \left( \frac{\rho}{\rho(1) }\right)^{3/4},\quad
 R_{4i} = 0,
\end{equation} 
and for $\rho_{\ast} \ge \rho$,
\begin{equation}
 R_{3i} = 0,\quad
 R_{4i} = \rho_{3i} (1) \left( \frac{\rho}{\rho(1) }\right) \alpha (\infty),
\end{equation} 
where $\alpha (\infty)$ is a constant defined by
\begin{equation}
 \alpha (\infty) := \int^{\infty}_{\beta} dt \left( \frac{t}{\beta} \right)^{1/2} 
 \exp{(-t + \beta)},
\end{equation}
which is evidently larger than the unity.
In order to minimize the difference between the junction model and the DD model,
we defined $R_{4i}$ so that $R_{4i}$ agrees with $\rho_{4i}$ at $\rho \to 0$.
By assuming that $R_{3i} = R_{4i}$ at $\rho = \rho_{\ast}$, the junction surface
$\rho_{\ast}$ is determined by
\begin{equation}
 \rho_{\ast} = \rho(1) \frac{1}{\alpha (\infty)^4}.
\end{equation}
Since the counterpart of the junction model corresponding with $\gamma_i$ of 
the DD model; the effective decay constant $\Gamma_i$ is given by
\begin{equation}
 \Gamma^2_i := \frac{\kappa^2}{3} \rho_{\ast},
\end{equation}
the relation between $\Gamma_i$ and $\gamma_i$ is given by
\begin{equation}
 \Gamma_i = \frac{1}{2 \beta \cdot \alpha(\infty)^2} \gamma_i. 
\end{equation}
Integrating the DD (\ref{integrationscalefactor}) under the junction model 
gives $N - N(0)$ as follows:
As for the righthand side of the decomposition of $(N - N(0))$ as 
\begin{equation}
 N - N(0) = \left(N - N_{\ast} \right) + \left(N_{\ast} - N(1) \right) 
 + \left( N(1)-N(0) \right),
\end{equation}
the first term is given by
\begin{align}
 N - N_{\ast} &= - 
 \frac{\rho(1)}{4 \left[ \rho_{3i}(1) \alpha(\infty) + \rho_{4r}(1) \right]}
 \ln{\frac{\rho}{\rho_{\ast}}}  \notag\\
 &- \frac{3 \rho_{3r} (1)}{16 \left[ \rho_{3i}(1) \alpha(\infty) + \rho_{4r}(1) \right]^2}
 4 \rho(1)^{5/4} 
 \left[ \left( \frac{1}{\rho} \right)^{1/4} - \left( \frac{1}{\rho_{\ast}} \right) ^{1/4} 
                 \right],
\end{align}
the second term is given by
\begin{align}
 N_{\ast}-N(1) &=- \frac{1}{3 \left[ \rho_{3i}(1) + \rho_{3r}(1) \right]}
 \rho(1)^{3/4} 4
 \left[ \rho^{1/4}_{\ast} - \rho(1)^{1/4} \right] \notag\\
 &+
 \frac{4 \rho_{4r}(1)}{9 \left[ \rho_{3i}(1) + \rho_{3r}(1) \right]^2}
 \rho(1)^{1/2} 2
 \left[ \rho^{1/2}_{\ast} - \rho(1)^{1/2} \right],
\end{align}
and the third term is given by
\begin{align}
 N(1) - N(0) &= - \frac{\rho(0)}{3 \left[ \rho_{3i}(0) + \rho_{3r}(0) \right]}
 \ln{\frac{\rho(1)}{\rho(0)}} \notag\\
 &+ \frac{4 \rho_{4r} (0)}{9 \left[ \rho_{3i}(0) + \rho_{3r}(0) \right]^2}
 \rho(0)^{2/3} 3
 \left[ \rho(1)^{1/3} - \rho(0)^{1/3} \right].
\end{align}
As for the subdominant energy components, we integrated after Taylor expanding 
the righthand side of Eq.(\ref{integrationscalefactor}).

Next we evaluate the difference of the Bardeen parameter between this junction model 
and the DD model whose $(N-N(0))$ cannot be integrated; $\Delta D (N-N(0))$ where 
we assume that $D=D(\rho, \rho(0))$.

\paragraph{Theorem 4.1}
Based on the junction model, the Bardeen parameter is evaluated as
\begin{equation}
 D (N-N(0)) = O(1) \frac{D \Gamma^2_i}{\Gamma^2_i} 
 + O(1) \frac{D \rho_{3i}(0)}{\rho (0)}
 + O(1) \frac{D \rho_{3r}(0)}{\rho (0)}
 + O(1) \frac{D \rho_{4r}(0)}{\rho (0)}, 
\end{equation}
where we consider the time region when $\rho_{3r}/\rho,\; \rho_{4r}/\rho \ll 1$.
The difference of the Bardeen parameter between in the junction model and in the DD model;
$\Delta D (N-N(0))$ is evaluated as
\begin{equation}
 \Delta D (N-N(0)) = O(1) \frac{D \Gamma^2_i}{\Gamma^2_i} 
 + O(1) \frac{D \rho_{3i}(0)}{\rho (0)}
 + O(1) \frac{D \rho_{3r}(0)}{\rho (0)}
 + O(1) \frac{D \rho_{4r}(0)}{\rho (0)}. 
\end{equation}

\vspace{0.5\baselineskip}
For the proof, please see appendix \ref{apdxproof4.1}.
According to this theorem, the junction model is not so good an approximation of the 
DD model as be expected. 
Although the junction model gives the right orders of the coefficients in the Bardeen
parameter, it does not give the right numerical factors of them.
But since in the DD model it is hopeless to obtain the analytic expression of the 
Bardeen parameter by integrating Eq.(\ref{integrationscalefactor}) and since we 
will be interested only in the evaluations of the orders of the coefficients which
will be performed by the exponent evaluation method \cite{Hamazaki2011}, we will
investigate the many step reheating by the junction method.

\section{dynamical property of the many step reheating}
\label{mainmanystep}

By the junction model, we investigate the many step reheating occurring after 
the inflation caused by the multiple slow rolling scalar fields. 
In this section and in the companion appendix \ref{apdxchangebase}, we put the assumption that
a certain initial fluctuation is not extremely large compared with the other initial fluctuations,
that is, the orders of all the initial fluctuations are not extremely different.
We assume that the metric junction surfaces $C_i$ ($i= 0, 1, \cdots, M$) are 
defined by 
\begin{equation}
 C_i := \rho (i) - \frac{3}{\kappa^2} \Gamma^2_i =0,
\end{equation}
where $\Gamma_i$ is the correspondent of the decay constant in the junction model
and is often modulated by another scalar field \cite{Sasaki1991}, \cite{Dvali2004}
which need not dominate the cosmic energy. 
From $\rho = \rho(0)$ to $\rho = \rho(1)$, all the scalar fields are in the oscillatory 
phase and are described as the $3$ fluids \cite{Kodama1996} \cite{Hamazaki1996}
\cite{Hamazaki2002} \cite{Hamazaki2004}.
At $\rho = \rho(i)$ ($i= 1, \cdots, M$), each $3$ fluid is changed into the $4$ fluid
one after another. 
Then from $\rho = \rho(M)$ to the present time, all the fluids are the $4$ fluids.
We write $\Delta N_i := N(i+1) - N(i)$ where $N(i)$ is the logarithm of the scale factor
at $\rho = \rho(i)$.
For the interval $\rho(0) \ge \rho \ge \rho(1)$, 
\begin{equation}
 \frac{\Gamma^2_1}{\Gamma^2_0} = \exp{[-3 \Delta N_0]}.
 \label{zeroblock}
\end{equation}
For the interval $\rho(i) \ge \rho \ge \rho(i+1)$ ($i= 1, \cdots, M-1$),
\begin{equation}
 \frac{\Gamma^2_{i+1}}{\Gamma^2_i} = s(i) \exp{[-3 \Delta N_i]} +
 \left(1- s(i) \right) \exp{[-4 \Delta N_i]},
\label{ithblock}  
\end{equation}
where $s(i)$ is the ratio of the energy density of the $3$ fluid to the total cosmic 
energy density 
at $\rho = \rho(i)$; $\rho_3(i) / \rho(i)$.
For $\rho(M) \ge \rho$,
\begin{equation}
 \frac{\kappa^2 \rho}{3 \Gamma^2_M} = \exp{[-4 (N - N(M))]}
 \label{Mthblock}
\end{equation}
where $\rho$, $N$ are the energy density, the logarithm of the scale factor at the present
time, respectively.
$s(i)$ ($i= 1, \cdots, M-1$) is pulled back to the initial time $\rho = \rho(0)$ as 
\begin{equation}
 s(i) = \frac{\rho_{3i}(0)}{\rho(i)} \exp{[-3 (N(i) - N(0))]},
 \label{sratio}
\end{equation}
where it can be considered that $s(0)=1$, $s(M)=0$.
Eq.(\ref{zeroblock}), Eq.(\ref{ithblock}), Eq.(\ref{Mthblock}) and Eq.(\ref{sratio})
are all the defining expressions of our many step reheating.
By these defining relations, our many step reheating is characterized by 
the degrees of freedom of the column $\rho_{3i}(0)$ ($i= 1, \cdots, M-1$) and
the degrees of freedom of the row $\Gamma^2_i$ ($i=0, 1, \cdots, M$) completely.
Next we present the formula which describes $D (N - N(0))$ in the form of the linear
combination of $D \rho_{3i}(0)$ ($i= 1, \cdots, M-1$) and 
$D \Gamma^2_i$ ($i=0, 1, \cdots, M$).
We interpret the perturbation operator $D$ as $D(C_0, C_1, \cdots, C_M, \rho)$.
Since in each interval $\rho(i) \ge \rho \ge \rho(i+1) $ ($i= 0, \cdots, M-1$)
we adopt $N$ as the evolution parameter describing all the fluid energy densities,
our treatment is rather implicit, but we can obtain the general expressions without
any subsidiary conditions about the sizes of the energy densities of the components.
The result is as follows: 
\begin{align}
 D (N - N(0)) &= (1+s_{M-1})(1+s_{M-2}) \cdots (1+s_1) \frac{1}{3} 
 \frac{D \Gamma^2_0}{\Gamma^2_0} \notag\\
 &+ \sum^{M}_{i=1} (1+s_{M-1})(1+s_{M-2}) \cdots (1+s_i) F_{i-1} 
 \frac{D \Gamma^2_i}{\Gamma^2_i} \notag\\
 &+ \sum^{M-1}_{i=1} (1+s_{M-1})(1+s_{M-2}) \cdots (1+s_{i+1}) 
 \left( - \frac{s_i}{3}\right) 
 \frac{D \rho_{3i}(0)}{\rho_{3i}(0)}. 
 \label{firstorderbardeen}
\end{align}
(For the derivation, please see the appendix \ref{apdxproofrecursive}.)
In order to understand Eq.(\ref{firstorderbardeen}), we introduce several notations.
As for 
\begin{equation}
 f_i = s(i) \exp{[-3\Delta N_i]} + (1-s(i)) \exp{[-4\Delta N_i]} 
\end{equation}
by regarding $s(i)$, $\Delta N_i$ as the variable $1$, the variable $2$, respectively,
we express the derivatives of $f_i$ with respect to them as
\begin{align}
 &(f_i)_1 = \frac{\partial}{\partial s(i)} (f_i), \quad
 (f_i)_2 = \frac{\partial}{\partial \Delta N_i} (f_i),\notag\\
 &(f_i)_{12} = \frac{\partial}{\partial s(i)}
     \frac{\partial}{\partial \Delta N_i} (f_i), \quad
 {\rm etc}     
\end{align}
We define $s_i$ by
\begin{equation}
 s_i = 3 \frac{(f_i)_1}{(f_i)_2} s(i).
\end{equation}
We define $F_{i-1}$ by
\begin{equation}
 F_{i-1} = (\delta_{i-1} -1) \frac{(f_{i-1})}{(f_{i-1})_2} \frac{1}{(g_{i-1})}
 s(i-1) \exp{[-3\Delta N_{i-1}]},
\end{equation}
where
\begin{equation}
 \delta_{i-1} = \frac{\rho_{3i}(0)}{\rho_{3 \; i-1}(0)},
\end{equation}
and
\begin{equation}
 g_{i-1} = (\delta_{i-1} -4) s(i-1) \exp{[-3\Delta N_{i-1}]} 
 - 4(1-s(i-1)) \exp{[-4\Delta N_{i-1}]}.
\end{equation}
When $\delta_{i-1} =1$, the $3$ fluid transferred into the $4$ fluid at $\rho = \rho(i)$
is vanishing.
Then $F_{i-1} =0$, which implies that the term of $D \Gamma^2_i$ in $D(N-N(0))$ 
disappears.
In order to interpret Eq.(\ref{firstorderbardeen}) by the exponent evaluation method 
\cite{Hamazaki2011}, we introduce the exponents $k_i$, $n_i$ and $m_i$ by
\begin{equation}
 s(i) =: 10^{- k_i},\quad \exp{[- \Delta N_i]} =: 10^{- n_i},\quad
 m_i := \min{\{ k_i, n_i \}}.
\end{equation}
In the same way as in the appendix \ref{apdxexponent}, from now on in this section,
for notational simplicity, we will simply write that $A \sim B \sim 10^{- \alpha}$ 
omitting the absolute value marks $|\cdots|$,
when we intend to imply $|A| \sim |B| \sim 10^{- \alpha}$ where $A$, $B$ are some physical quantities 
and $\alpha$ is the positive number.
In this way, we are only interested in the sizes of the physical quantities, while we neglect the signs of 
the physical quantities.
According to appendix \ref{apdxexponent}, we can evaluate
\begin{align}
 &s_i \sim 10^{- k_i + m_i}, \quad
 1+s_i \sim 10^{- n_i + m_i},\notag\\
 &F_i \sim 10^{- k_i + m_i} \times (\delta_i - 1) \sim (\delta_i - 1) s_i, 
\end{align}
which gives
\begin{align}
 {\rm for}\; k_i &> n_i,\quad s_i \ll 1, \quad 1+ s_i \sim 1 \notag\\
 {\rm for}\; k_i &= n_i,\quad s_i \sim 1,\quad 1+ s_i \sim 1 \notag\\
 {\rm for}\; k_i &< n_i,\quad s_i \sim 1,\quad 1+ s_i \ll 1. 
\end{align}
From the above, we can see that $s_i$ and $1+s_i$ are in the reciprocal relation,
that is, when the one is small, the other becomes large.
By seeing Eq.(\ref{firstorderbardeen}) in the viewpoint of this fact, we can see 
the following remarkable property.

Now we consider the following case, which will be called the $l$ case from now on.
This case is defined by
$k_l < n_l$, that is $s_l \sim 1$, $1+s_l \ll 1$ and
$k_i > n_i$ ($i= l+1, \cdots, M-1$), that is $s_i \ll 1$, $1+s_i \sim 1$.
In the $l$ case, in $D (N - N(0))$, only the term of $D \rho_{3l}(0) /\rho_{3l}(0)$
and the term of $D \Gamma^2_{l+1} / \Gamma^2_{l+1}$ do remain.
Since $1 + s_l \ll 1$, the coefficients of the previously generated perturbations such as
$D \rho_{3i}(0) /\rho_{3i}(0)$ ($i=1, 2, \cdots, l-1$) and
$D \Gamma^2_{i} / \Gamma^2_{i}$ ($i= 0, 1, \cdots, l$) become negligibly small.
In addition, since $F_i \sim s_i \ll 1$ ($i=l+1, \cdots, M-1$), 
the terms of $D \rho_{3i}(0) /\rho_{3i}(0)$ ($i=l+1, \cdots, M-1$) and
the terms of $D \Gamma^2_{i} / \Gamma^2_{i}$ ($i= l+2, \cdots, M$) are not effective.
These facts can be interpreted in the physical way.
Our many step reheating has the property of oblivion by which we mean the property that
the component governing the cosmic energy density governs the adiabatic perturbation 
represented by the Bardeen parameter and the effects of 
the perturbations of all the components diluted by the cosmic expansion become negligibly small. 
As shown in appendix \ref{apdxoblivion}, the property of oblivion is properly possessed by 
multiple fluids system and it is not by other systems such as
the multiple slow rolling scalar fields system and the system appearing in the ordinary inflationary 
cosmologies constructed by connecting
the slow rolling scalar fields phase and the perfect fluids phase by the metric junction.
In the system consisting of the $3$ fluid and the $4$ fluid, since the energy density of
the $n$ fluid is diluted in proportion of the reciprocal of $a^n$ by the cosmic expansion,
the ratio occupied by the energy density of the $3$ fluid to the total energy density 
becomes predominant.
In the $l$ case, in the $l$ step ($\rho(l) \ge \rho \ge \rho(l+1)$), the energy dominance
of the $3$ fluid is attained and in the following steps, the cosmic expansion does not
proceed so that the $4$ fluid originating from the $3$ fluid in the $l$ step remains 
dominant.   
That $D \rho_{3l}(0) /\rho_{3l}(0)$ is effective corresponds with the curvaton mechanism 
\cite{Lyth2002} \cite{Ichikawa2008} \cite{Sasaki2006} \cite{Assadullahi2007} \cite{Suyama2011}.
In addition, since at $\rho = 3 \Gamma^2_{l+1} / \kappa^2$
the $3$ fluid to have dominated the cosmic energy in the $l$ step
is changed into the $4$ fluid which remains to govern the cosmic energy until the present 
time, the term of $D \Gamma^2_{l+1} / \Gamma^2_{l+1}$ is effective.
By the $3$ fluid dominating the cosmic energy in the $l$ step, 
the $4$ fluid previously generated from the $3$ fluid is diluted so much that
all the perturbations originating from
$D \rho_{3i}(0) /\rho_{3i}(0)$ ($i=1, 2, \cdots, l-1$) and
$D \Gamma^2_{i} / \Gamma^2_{i}$ ($i= 0, 1, \cdots, l$) fall into oblivion.
Therefore the Bardeen parameter in the $l$ case becomes
\begin{equation}
 D (N - N(0))_l = O(1) \frac{D \Gamma^2_{l+1}}{\Gamma^2_{l+1}}
 + O(1) \frac{D \rho_{3l}(0)}{\rho_{3l}(0)}.
\label{lcasefirst} 
\end{equation}

The property of oblivion also governs the second order perturbation of the 
many step reheating.
Please notice
\begin{equation}
 \frac{D s(i)}{s(i)} = \frac{D \rho_{3i}(0)}{\rho_{3i}(0)}
 - \frac{D \Gamma^2_i}{\Gamma^2_i}
 - 3 D (N(i) - N(0)),
\end{equation}
where the third term of the righthand side is given by
\begin{align}
 &D (N(i) - N(0)) = (1+s_{i-1})(1+s_{i-2}) \cdots (1+s_1) \frac{1}{3} 
 \frac{D \Gamma^2_0}{\Gamma^2_0} \notag\\
 &+ \sum^{i-1}_{j=1} (1+s_{i-1})(1+s_{i-2}) \cdots (1+s_j) F_{j-1} 
 \frac{D \Gamma^2_j}{\Gamma^2_j} \notag\\
 & + \frac{(f_{i-1})}{(f_{i-1})_2} \frac{D \Gamma^2_i}{\Gamma^2_i}
 + \sum^{i-1}_{j=1} (1+s_{i-1})(1+s_{i-2}) \cdots (1+s_{j+1}) 
 \left( - \frac{s_j}{3}\right) 
 \frac{D \rho_{3j}(0)}{\rho_{3j}(0)},
\end{align}
and
\begin{align}
 &D \Delta N_i = s_i (1+s_{i-1})(1+s_{i-2}) \cdots (1+s_1) \frac{1}{3} 
 \frac{D \Gamma^2_0}{\Gamma^2_0} \notag\\
 &+ \sum^{i-1}_{j=1} s_i (1+s_{i-1})(1+s_{i-2}) \cdots (1+s_j) F_{j-1} 
 \frac{D \Gamma^2_j}{\Gamma^2_j} \notag\\
 & + \left[ - \frac{1}{s(i)-4} + s_i F_{i-1} \right]
 \frac{D \Gamma^2_i}{\Gamma^2_i}
 + \frac{(f_{i})}{(f_{i})_2} \frac{D \Gamma^2_{i+1}}{\Gamma^2_{i+1}} \notag\\
 &+ \sum^{i-1}_{j=1} s_{i} (1+s_{i-1})(1+s_{i-2}) \cdots (1+s_{j+1}) 
 \left( - \frac{s_j}{3}\right) 
 \frac{D \rho_{3j}(0)}{\rho_{3j}(0)}
 - \frac{s_i}{3} \frac{D \rho_{3i}(0)}{\rho_{3i}(0)}. 
\end{align}
Please also notice the exponent evaluation as
\begin{equation}
 (s_i)_1 s(i) \sim (s_i)_2 \sim (F_i)_1 s(i) \sim (F_i)_2 \sim s_i (1+s_i),
\label{evaluatesecond} 
\end{equation}
and
\begin{equation}
 \frac{d}{d \delta_i} F_i \sim s_i.
 \label{evaluatesecondtwo}
\end{equation}
For the derivation, please see appendix \ref{apdxexponent}.
We consider operating the perturbation operator $D$ on Eq.(\ref{firstorderbardeen})
in order to obtain $D^2 (N-N(0))$. 
Since from Eq.(\ref{evaluatesecond}), $s_i$, $1+s_i$ and $F_i$ operated the perturbation 
operator $D$ are of order of $s_i (1+s_i) \ll 1$, they are not effective in $D^2 (N-N(0))$.
But from Eq.(\ref{evaluatesecondtwo}), only $D \delta_l$ given by
\begin{equation}
 D \delta_l = \delta_l \left[ \frac{D \rho_{3 \; l+1}(0)}{\rho_{3 \; l+1}(0)} 
 - \frac{D \rho_{3l}(0)}{\rho_{3l}(0)} \right]
\end{equation}
is effective.
Then for the $l$ case, the second order perturbation of the Bardeen parameter $D^2 (N-N(0))$
is evaluated as
\begin{equation}
 D^2 (N - N(0))_l = O(1) D \left( \frac{D \Gamma^2_{l+1}}{\Gamma^2_{l+1}} \right)
 + O(1) D \delta_l \cdot \frac{D \Gamma^2_{l+1}}{\Gamma^2_{l+1}}
 + O(1) D \left( \frac{D \rho_{3l}(0)}{\rho_{3l}(0)} \right).
\end{equation}
Therefore in order to evaluate the Bardeen parameters $D^n (N-N(0))$ ($n=1,2$), 
it is sufficient only to investigate the step the nearest to the present time 
when the $3$ fluid dominates the cosmic energy, that is the $l$ step
in almost all the cases, as long as any specific
$D \Gamma^2_i / \Gamma^2_i$, $D \rho_{3i}(0) / \rho_{3i}(0)$ are not extremely large.

The property of oblivion which brings about the large simplification to the calculation
of the Bardeen parameters in the many step reheating can also be seen in the results of 
the researches of the curvaton scenario \cite{Lyth2002} \cite{Ichikawa2008} \cite{Sasaki2006} 
\cite{Assadullahi2007} \cite{Suyama2011}, although the authors have not given a name to 
this property.
In particular, in the paper \cite{Suyama2011} treating the system with the two curvaton decays,
by their rather complicated calculation the appearance of the property of oblivion can be found.
We succeeded in proving that the property of oblivion appears in arbitrarily many step reheating
by devising the way of giving the initial data and by transparent recursive calculation.
By this we can clarify which of the entropic information of the multiple slow rolling scalar 
fields system in the inflationary period has influence on the Bardeen parameter 
at the present time through the many step reheating.
 
So far we determined the time evolution of the Bardeen parameters in the many step reheating 
by solving the recursive relations connecting one energy transfer with next energy transfer.
In the appendix \ref{apdxchangebase}, the same problem will be solved by the formula of the 
change of the base formulated in the section \ref{mainchangebase}.
There the comparison of our formulation with the papers of $\delta N$ formalism \cite{Sasaki2006}, 
the separate universe approach \cite{Wands2000}
will be discussed in order to meet the request of the reviewer.

\section{Summary and Discussion}

We will summarize the flow of the present paper.
In section $2$, we give the definition of the gauge invariant perturbation variable in an arbitrary
high order and in the arbitrary background spacetime and the proof of its gauge invariance.
It is our answer to the problem presented by Dr. Nakamura in the paper \cite{Nakamura2014}.
The purpose of the section $2$ is to clarify how to write the solutions 
which contains the plural times in general, of the physical laws in the manifestly gauge invariant manner.
In this section, we extend the definition of the background/scalar like object from the 
previous single time case \cite{Hamazaki2008.2} \cite{Hamazaki2011} to the many time case
and the gauge invariance of the many time background/scalar like object is proven.
The solutions can be interpreted as the interrelations between the physical quantities 
at plural times, most simply, at the initial time and at the final time.
Therefore the solutions can be shown to be written in the manifestly gauge invariant 
manner, concretely speaking, in the form in which many time
background/scalar like object written in terms of the single time background/scalar like
objects at the plural times is vanishing.
In general, the background/scalar like object has the scalar like object as the base,
and the background/scalar like object as the true value
and can be interpreted as the fluctuation of the true value in the time slice in which 
the base does not fluctuate.
While in the many time background/scalar like object with the $n$ times, its base is 
thought ordinarily to be the set
of the $n$ single time scalar like objects at the $n$ different times,
it is possible and useful in the viewpoint of the applications to take as the base 
the $n$ independent many time scalar like objects depending on the plural times,
which corresponds with specifying the $n$ different time slices at the $n$ times 
simultaneously.
In the paper \cite{Malik2009}, Malik and Wands suggested another way of defining the gauge invariant
perturbation variable.
We show that our definition and the definition in the fashion of Malik, Wands are equivalent
in spite of the apparent difference. 
The difference between the both is caused by the difference of way of the parametrization of the gauge
vector. 

In the section $3$, we give the relation connecting the background/scalar like objects 
with arbitrary two different sets of the bases in case of general many time background/scalar like objects,
since it is desirable to be able to freely change the time slices of the initial time 
and the final time in which the perturbations are considered.
It is called the formula on transform of the bases. 
In the single time case also, it is easy to derive the formula on transform of the bases 
in such case.
As the application of the formula of transform on the bases, we consider the multiple 
slow rolling scalar fields system.
In this system, the $\tau$ function is used as the evolution parameter 
\cite{Hamazaki2008.2} \cite{Hamazaki2011}, and in its expression of the Bardeen parameter
the perturbations of the $\tau$ function not related with the observation are contained.
In order to kick the perturbations of the $\tau$ function out of the the Bardeen parameter,
the formula on the transform of the bases is successfully used.

As a successful application of the discussion in the sections $2$, $3$, that is of the expression of 
the solution in the viewpoint of the many time background/scalar like object, in the sections $4$, $5$
we consider the many step reheating which contains an arbitrary large number of energy transfer
processes \cite{Elliston2014} \cite{Meyers2014} \cite{Leung2012}.
The solution of the many step reheating based on the junction model can be interpreted 
in terms of the many time background/scalar like object with the initial time, the final time
whose time slices can be freely chosen, and with the many times specifying 
the junction surfaces \cite{Hamazaki2011} in which the $3$ fluids are transferred 
into the $4$ fluids, whose time slices cannot be changed freely.
In the section $5$, we consider the many step reheating as the junction model in which the transfers 
from the $3$ fluid to the $4$ fluid are described as the metric junctions 
\cite{Israel1966} \cite{Mukohyama2000} \cite{Hamazaki2011} \cite{Elliston2014} \cite{Meyers2014}
\cite{Sasaki2006} \cite{Assadullahi2007} \cite{Suyama2011},    
while the reheating is described by the system of the differential equations with 
the decay term of the fluids \cite{Hamazaki1996}.
In the section $4$, we discuss the legitimacy of the junction model of the reheating by solving the system 
of the differential equations with the decay term and the junction model
using the energy density of the total system $\rho$ as
the evolution parameter \cite{Hamazaki2011}, and by evaluating the difference between 
both.
The replacement of the effect of the decay term with the metric junction can
simplify the solution remarkably.

In the section $5$, we use the junction model but we use the scale factor $a$
as the evolution parameter because we can write down the solution without assuming
the conditions as for the energy ratios between the components, while the expressions of 
the solution are implicit.
We propose the useful parametrization treating the many step reheating, which is the combination
of the expressions describing the evolutions between one energy transfer and the next energy 
transfer and the expressions solving from the initial time of the many step reheating, that is,
the end of the inflation, the energy density of the $3$ fluid at the initial time of this step.
The Bardeen parameter is given by solving this combination recursively.
We interpret the solution by the exponent evaluation method \cite{Hamazaki2011} in which
attention is paid only to the orders of the sizes of the physical quantities.
In result, we find the remarkable property, the property of oblivion, which implies that compared with
the perturbations of the processes generating the perturbation effectively the effects of the perturbations 
existing previously become negligibly small.
The property of oblivion is realized in the system in which the perturbation of the component 
governing the cosmic energy contributes the Bardeen parameter mainly.
This property holds in the multiple fluid system and does not hold in the multiple slow rolling scalar fields system.
In the junction models also, the property of oblivion is not self-evident since as shown in the appendix 
\ref{apdxoblivion} the property of oblivion does not hold in an ordinary history appearing in the standard inflationary
cosmologies in which the slow rolling phase caused by the multiple scalar fields and the fluid phase 
in which the multiple scalar fields have been transformed into the dust-like/radiation fluids,
are combined by the metric junction.
But our many step reheating satisfies the property of oblivion.
By the property of the oblivion, in almost all cases in the many step reheating
it is sufficient to pay attention only to the perturbations related with the cosmic energy component 
governing the present universe and to take into account only the effects of 
the prominent perturbation generation mechanisms such as the curvaton \cite{Lyth2002} 
\cite{Ichikawa2008} \cite{Sasaki2006} \cite{Assadullahi2007} \cite{Suyama2011}   
and the modulated reheating \cite{Sasaki1991} \cite{Dvali2004} 
the nearest to the present time.
That is, in the present Bardeen parameter, only the fluctuation of the $3$ fluid governing the cosmic energy finally and
the fluctuation of the decay constant with which this $3$ fluid decays into the $4$ fluid are dominant.

So far, we consider the gauge invariant perturbation variables order by order with respect to the 
perturbation parameter $\lambda$.
In contrast, there exists an approach where the gauge invariant variables full with respect to the finite $\lambda$
are considered based on the canonical theory of the constrained system \cite{Giesel2010}.
Although this approach is interesting, the gauge invariance in the paper \cite{Giesel2010} means the gauge invariance
in the canonical theory of the constrained system while our gauge invariance is the 
invariance for the gauge transformation generating the change of the point identification between the perturbed universe
$M_{\lambda}$ and the unperturbed universe $M_{\lambda=0}$, that is the gauge transformation of the second kind.
Therefore the above two gauge invariance are not the same completely from the conceptual point of view.
In addition, the original general relativity is the time reparametrization invariant, the evolutions of the full order gauge
invariant variables become self-evident.
The paper \cite{Giesel2010} changes the original general relativity in order to  obtain the non-self-evident evolutions 
of their full order gauge invariant variables and the deviation of the thoery \cite{Giesel2010} from the original
general relativity is finite, although it is small in the practical case.

\section*{Acknowledgments}

The author would like to thank Professors H. Kodama and J. Yokoyama
for continuous encouragements.
The author would like to thank Dr. K.Nakamura for fruitful conversation and his writing the paper \cite{Nakamura2014}
from which  he learned that in the definition of the gauge invariant variable the Lie derivative can 
be effectively used. 
The author would like to thank the referee for pointing his interest to the interesting paper \cite{Giesel2010}.
The author would like to thank reviewers for giving suggestions helping him to make this paper self-contained.
The author would like to thank the other referee for evaluating his work highly and giving the warm encouraging comments.

\appendix

%koko

%% Equation numbering %%
% koko de Appendix no siki banngou wo (A,1) ni suru

\catcode`\@=11

\@addtoreset{equation}{section}   % Makes \section reset 'equation' counter.
\def\theequation{\Alph{section}.\arabic{equation}}

%koko

\section{On the derivation of the gauge invariant perturbation variable by K.Nakamura}
\label{apdxnakamura}

We discuss the derivation of the gauge invariant perturbation variable in the general spacetime and for arbitrarily
high orders by K.Nakamura based on the paper
\cite{Nakamura2014}.
We would like the reader to read this appendix after reading the sections $2$.
K.Nakamura used another parametrization of the gauge vector than ours and that of Malik, Wands, although 
the correspondence between the Nakamura's gauge vector and those of the others can be calculated by the
Baker Campbell Hausdorff formula.
In order to skip the explanation of the further notational complexity, 
we would like to explain Nakamura's derivation of the gauge invariant perturbation variable by using our parametrization 
of the gauge vector introduced in the first half of the section $2$.
According to Theorem $GMW.3$, the gauge transformed variable ${d^k}/{d \lambda^k} \cdot X(\mu = 1) |_{\lambda = 0}$
($k=0,1,2,\cdots$) where $X$ is an arbitrary tensor is given by
\begin{equation}
 \frac{d^k}{d \lambda^k} X(\mu = 1) \Bigg|_{\lambda = 0} =
 \left[ \frac{d}{d \lambda} + L\left( T_{\ast} \right)   \right]^k  X(\mu = 0) \Bigg|_{\lambda = 0},
\end{equation}
where the relation between the vector generating the gauge transformation $T$ and $T_{\ast}$ is given by 
Theorem $GMW.3$.
We conjecture that 
$[ {d}/{d \lambda} - L\left( V \right)  ]^k  X|_{\lambda = 0}$ with $T_{\ast}$ replaced with $-V$
in the expression in the right hand side of the above equation, is gauge invariant.
Then we define $Q_k$ by
\begin{equation}
 Q_k := \left[ \frac{d}{d \lambda} - L\left( V \right)   \right]^k  X \Bigg|_{\lambda = 0}
+  L\left( \frac{d^{k-1} V}{d \lambda^{k-1}}\right)  X \Bigg|_{\lambda = 0} ,
\end{equation}
where the right hand side is written by
$ {d^{l} V}/{d \lambda^{l}}|_{\lambda = 0} $ ($l=0,1,2,\cdots,k-2$) only.
By the lengthy direct calculation, we derive that there exists $P_{k-1}$ satisfying
\begin{equation}
 Q_k (\mu=1) - Q_k (\mu=0) = L(P_{k-1}) X |_{\lambda=0}.
\label{lengthy}
\end{equation}
In the paper \cite{Nakamura2014}, K.Nakamura confirmed the existence of $P_0$, $P_1$, $P_2$, $P_3$ by the direct
calculation and for the general natural number $k$, the existence of $P_k$ satisfying (\ref{lengthy})
was assumed as Conjecture $4.1$.
By using $ {d^{k-1} V}/{d \lambda^{k-1}} |_{\lambda = 0}$ whose existence is assumed in Conjecture $3.1$
in the paper \cite{Nakamura2014} having the gauge transformation property:
\begin{equation}
 \frac{d^{k-1}}{d \lambda^{k-1}} V(\mu = 1) \Bigg|_{\lambda = 0} -
\frac{d^{k-1}}{d \lambda^{k-1}} V(\mu = 0) \Bigg|_{\lambda = 0} = P_{k-1},
\end{equation}
we can conclude that the following variable defined by
\begin{equation}
   Q_k -   L\left( \frac{d^{k-1} V}{d \lambda^{k-1}}\right)  X \Bigg|_{\lambda = 0} =
  \left[ \frac{d}{d \lambda} - L\left( V \right)   \right]^k  X \Bigg|_{\lambda = 0}                          
\end{equation}
is gauge invariant.
If we perform the Nakamura like derivation, the derivations of equations (\ref{lengthy}) for an arbitrary natural 
number $k$ are hopelessly difficult.

By using the much simpler calculation in the section $2$, we try to derive the concrete expression of $P_{k-1}$ 
in the equation (\ref{lengthy}).
The infinitesimal gauge transformations of the perturbations of the gauge vector    
 $ {d}/{d \mu} \cdot {d^{k}/}{d \lambda^{k} } \cdot V  $ for $k=0,1,2,\cdots$ are given by
\begin{equation}
 \frac{d}{d \mu}  \frac{d^{k-1}}{d \lambda^{k-1}} V = \frac{d^k}{d \lambda^k} T
 + \sum^{k-1}_{l=0} {_{k-1} C_{l} } 
 \left[ \frac{d^l}{d \lambda^l} T,
 \frac{d^{k-1-l}}{d \lambda^{k-1-l}} V
\right].
\end{equation}
Using such above expressions, we obtain the concrete, rather complicated expressions of $P_{k-1} |_{\lambda =0}$
in the following  form as
\begin{equation}
P_{k-1} |_{\lambda =0} =
 \sum^{\infty}_{l=1} \frac{1}{l!}
 \frac{d^l}{d \mu^l}  \frac{d^{k-1}}{d \lambda^{k-1}} V \Bigg|_{\lambda=0},
\end{equation}
although the right hand side contains the finite terms only for finite $k$ because of $T(\lambda =0) =0$.
In this way, the Nakamura's calculations are rather complicated because K.Nakamura considers the gauge
transformation between the physical quantities with the finitely separated gauge parameters such as 
$\mu=0$ and $\mu=1$.
If we only consider the infinitesimally separated gauge transformation, that is $d X / d \mu$ as in our section
$2$, the calculation becomes drastically simple.

In the section $5$ in the paper \cite{Nakamura2014},
K.Nakamura discussed the construction of the gauge vector satisfying the appropriate gauge transformation property.
But since he does not have the concept of the base, it is difficult to understand its physical meaning.
It is however useful to try to understand the meaning of the equation ($81$) in the paper \cite{Nakamura2014} 
by our words. 
In our method, as explained in the appendix \ref{apdxgeneralbase}, we iteratively solve the equation (\ref{oneafteranother})
obtained by the $d / d \lambda$ differentiation of  equation (\ref{definitiongaugevectorbybase}),
with respect to $d^k V / d \lambda^k |_{\lambda =0}$ ($k=0,1,2,\cdots$).
If K.Nakamura had had the concept of the base, he would have used the following equation and solved it
with respect to  $d^k V / d \lambda^k |_{\lambda =0}$ ($k=0,1,2,\cdots$) iteratively:
\begin{equation}
 L \left( \frac{d^k}{d \lambda^k} V \right) S \Bigg|_{\lambda =0} =
 \left[ \frac{d}{d \lambda} - L(V)  \right]^{k+1} S \Bigg|_{\lambda =0} +
 L \left( \frac{d^k}{d \lambda^k} V \right) S \Bigg|_{\lambda =0}, 
\label{nakamuragauge}
\end{equation}
whose right hand side does not contain  $d^k V / d \lambda^k |_{\lambda =0}$ and corresponds to 
the K.Nakamura notation ${^{(k)} \hat{H}_{ab}}$.
While we use the equation (\ref{oneafteranother}) obtained by the $d / d \lambda$ differentiation of  equation(\ref{definitiongaugevectorbybase}),
K.Nakamura uses the equation (\ref{nakamuragauge})
obtained by the $[d / d \lambda - L(V)]$ differentiation of  equation (\ref{oneafteranother}).
Therefore compared with our method, the Nakamura method generates the extra terms.
But both our method and the Nakamura method are equivalent.

It would be useful that when we have some sequence of the gauge invariant perturbation variables,
we can make other sequences of the gauge invariant perturbation variables as stated in the following way. 
We still consider the general spacetime.

\paragraph{Theorem $GA.1$}
Assuming that we have all the coefficients of the $\lambda$ expansion of the gauge vector 
$d^k V / d \lambda^k |_{\lambda =0}$ ($k=0, 1, 2, \cdots$) which make the perturbation variables defined by
\begin{equation}
 \left[ \frac{d}{d \lambda} - L\left( V \right)   \right]^k  S \Bigg|_{\lambda = 0}   \quad(k =1, 2, \cdots),
\label{originalgauge}
\end{equation}
where $S$ is some invertible set of the physical quantities, are gauge invariant, for an arbitrary physical 
quantity $X$, the perturbation variables defined by  
\begin{equation}
 \left[ \frac{d}{d \lambda} - L\left( V \right)   \right]^k  X \Bigg|_{\lambda = 0}   \quad(k =1, 2, \cdots)
\end{equation}
are also gauge invariant.
\vspace{0.5\baselineskip}

\paragraph{Proof of Theorem $GA.1$}

From (\ref {originalgauge}), using the correspondence $V \to \tilde{V}$ in Theorem $G6.3$ we obtain
\begin{equation}
 \frac{d}{d \mu} \exp{[- L(\tilde{V})]} S =0.
\end{equation}
Then by the same discussion as in the proof of Theorem $G6.1$, $\tilde{V}$ is the gauge vector of the
second type.
Therefore by using the correspondence $\tilde{V} \to V$ in Theorem $G6.3$, for all the nonnegative integers $k$,
the perturbation variables as
\begin{equation}
  \left[ \frac{d}{d \lambda} - L\left( V \right)   \right]^k  X \Bigg|_{\lambda = 0} =
\frac{d^k}{d \lambda^k} \left[ \exp{[- L(\tilde{V})]} S \right] \Bigg|_{\lambda =0}
\end{equation}
are also gauge invariant. 
\vspace{0.5\baselineskip}

\section{the concept of the base of the gauge invariant perturbation variable in the general spacetime}
\label{apdxgeneralbase}

One of the referees said that our definition of the gauge invariant perturbation variable cannot be used
in the case of the general spacetime and is confined in the zeroth order of the gradient expansion.
Then in this appendix, we will show that also in the general spacetime, our concept of the base is 
effective in the construction of the gauge invariant perturbation variable without using the gradient 
expansion.   

We consider the perturbed homogeneous isotropic universe.
By using the notation in the textbook \cite{Kodama1991}, the gauge vector is parameterized 
as $V^{\mu} = (T, L^i)$ and the metric is parametrized as
\begin{align}
 g_{00} &= - (1 + 2 \alpha),\notag\\
 g_{0j} &= - a \beta_j, \notag\\
 g_{ij} &= a^2 (\gamma_{ij} + 2 h_{ij}),
\end{align}
where $\gamma_{ij}$ is the metric of the three dimensional constant curvature space.
Three dimensional vectors $v = L, \beta$ are decomposed as
\begin{equation}
 v_j = D_j v_L + v_{Tj}, \quad D_j v^j_T =0, 
\end{equation}
where $D_j$ is the covariant derivative associated with $\gamma_{ij}$.
Three dimensional tensor $t = h$ is decomposed as
\begin{align}
 t_{jk} &= \gamma_{jk} t_L + (D_j D_k -\gamma_{jk} D^2/3) t_T + (D_j t_{Tk} + D_k t_{Tj})/2 + t_{TTjk},
\notag\\
 D_j t^j_T &= D_l t_{TTj}^l = t_{TTl}^l =0.
\end{align}
The physical quantities $X$ given by
\begin{equation}
 X = T, L_L, L_{Ti}, \alpha, \beta_{L}, \beta_{Ti}, h_L, h_T, h_{Ti}, h_{TTij}, 
\end{equation}
are expanded with respect to the perturbation parameter $\lambda$ as
\begin{equation}
 X = \sum^{\infty}_{k=1} \frac{\lambda^k}{k!} \frac{d^k X}{d \lambda^k} \Big|_{\lambda =0}.
\end{equation}

The values of the Lie derivatives induced by the gauge vector $V$ at $\lambda =0$ are 
calculated in the following way;
\begin{align}
 L(V) \alpha |_{\lambda =0} &=  \dot{T},    \notag\\
 L(V) \beta_L |_{\lambda =0} &= - a \dot{L}_L + \frac{T}{a}, \notag\\
 L(V) h_L |_{\lambda =0} &= \frac{1}{3} D^2 L_L +\frac{\dot{a}}{a} T, \notag\\
 L(V) h_T |_{\lambda =0} &= L_L,  \notag\\
 L(V) \beta_{Tj} |_{\lambda =0} &= - a\dot{L}_{Tj}, \notag\\
 L(V) h_{Tj} |_{\lambda =0} &=  L_{Tj},  \notag\\
 L(V) h_{TTij} |_{\lambda =0} &= 0,     
\label{Liederivativezerolambda}           
\end{align}
where all quantities in the right hand side are evaluated at $\lambda =0$, although $|_{\lambda =0}$'s 
are omitted.
The gauge vector $V$ is obtained as the unique solution to the equation with 
the invertible base $S$ given by
\begin{equation}
 D(S) S:= \left[ \frac{d}{d \lambda} -L(V) \right] S =0.
\label{definitiongaugevectorbybase}
\end{equation}
Practically the gauge vector $V$ is determined by recursively solving with respect to 
 $d^k V / d \lambda^k |_{\lambda=0}$ ($k=0,1,2, \cdots$) the following equations;
\begin{equation}
 L \left( \frac{d^k}{d \lambda^k} V \right) S \Bigg|_{\lambda =0} =
 \frac{d^{k+1}}{d \lambda^{k+1}} S \Bigg|_{\lambda =0}
- \sum^{k-1}_{l=0} {_k C_l} \; L \left( \frac{d^l}{d \lambda^l} V \right)  
   \frac{d^{k-l}}{d \lambda^{k-l}}  S \Bigg|_{\lambda =0},
\label{oneafteranother}
\end{equation}
where we notice in the right hand side $d^k V / d \lambda^k |_{\lambda=0}$ does not appear.
Being able to define the gauge vector $V$ uniquely, that is the base $S$ is invertible, is equivalent to 
being able to solve the following problem:

\paragraph{Problem $A$}
For an arbitrary quantity $A$, can we uniquely solve the following equation with respect to the vector $V$?
\begin{equation}
 L(V) S |_{\lambda =0} =A,
\end{equation}
whose left hand side can be calculated by (\ref {Liederivativezerolambda} ) .
\vspace{0.5\baselineskip}

We cannot adopt $h_{TTij}$ as one of the base, since $L(V) h_{TTij} |_{\lambda =0}$ is vanishing.
Whatever gauge vector we choose, we cannot satisfy (\ref {oneafteranother}).
We cannot  put $S$ as the base when
$L(V) S |_{\lambda =0}$ contains only the time derivatives such as $\dot{T}$, $\dot{L}_L$, $\dot{L}_{Tj}$.
This is because integrating the time derivatives induces the integration constants spoiling the uniqueness of 
the gauge vector. 

We will present some successful invertible bases.
Let the base $S_1$ be $S_1= (\sigma_{gL}, h_T, h_{Ti})$, in other words $S_1= (\beta_L, h_T, h_{Ti})$.
Under this base, the linear gauge invariant perturbation variables are given by
\begin{align}
 D(S_1) \alpha &= \frac{d \alpha}{d \lambda} - \left( \frac{d \sigma_{gL}}{d \lambda} \right)^{\cdot},
\notag\\
 D(S_1) \beta_L &= 0, \notag\\
 D(S_1) h_L &= \frac{d \mathcal{R}}{d \lambda} - \frac{\dot{a}}{a} \frac{d \sigma_{gL}}{d \lambda},\notag\\
 D(S_1) h_T &= 0, \notag\\
 D(S_1) \beta_{Ti} &=  \frac{d \beta_{Ti}}{d \lambda} + a \frac{ d \dot{h}_{Ti}}{d \lambda} , \notag\\
 D(S_1) h_{Ti} &= 0, \notag\\
 D(S_1) h_{TTij} &= \frac{d h_{TTij}}{d \lambda}, 
\end{align}
where all quantities are evaluated at $\lambda =0$ though all $|_{\lambda =0}$'s are omitted and where
\begin{align}
 \sigma_{gL} &:= a \beta_L + a^2 \dot{h}_T, \notag\\
 \mathcal{R} &:= h_L - \frac{1}{3} D^2 h_T.
\end{align}
Under the base $S_1$,
the independent higher order gauge invariant perturbation variables are given by
\begin{equation}
 D(S_1)^k X, \quad X=\alpha, h_L, \beta_{Ti}, h_{TTij},
\end{equation}
for an arbitrary natural number $k$.

Let the base $S_2$ be $S_2= (\mathcal{R}, h_T, h_{Ti})$, in other words $S_2= (h_L, h_T, h_{Ti})$.
Under this base, the linear gauge invariant perturbation variables are given by 
\begin{align}
 D(S_2) \alpha &= \frac{d \alpha}{d \lambda} - \left( \frac{a}{\dot{a}} \frac{d \mathcal{R}}{d \lambda} \right)^{\cdot},
\notag\\
 D(S_2) \beta_L &= - \frac{1}{\dot{a}} 
\left( \frac{d \mathcal{R}}{d \lambda} - \frac{\dot{a}}{a} \frac{d \sigma_{gL}}{d \lambda}  \right) , \notag\\
 D(S_2) h_L &= 0,\notag\\
 D(S_2) h_T &= 0, \notag\\
 D(S_2) \beta_{Ti} &=  \frac{d \beta_{Ti}}{d \lambda} + a \frac{ d \dot{h}_{Ti}}{d \lambda} , \notag\\
 D(S_2) h_{Ti} &= 0, \notag\\
 D(S_2) h_{TTij} &= \frac{d h_{TTij}}{d \lambda}, 
\end{align}
all quantities are evaluated at $\lambda =0$ though all $|_{\lambda =0}$'s are omitted.
Under the base $S_2$,
the independent higher order gauge invariant perturbation variables are given by
\begin{equation}
 D(S_2)^k X, \quad X=\alpha, \beta_L, \beta_{Ti}, h_{TTij},
\end{equation}
for an arbitrary natural number $k$.

As another invertible base, we can adopt $S_3 = (s, h_T, h_{Ti})$ where $s$ is the scalar quantity such as
the energy density $\rho$.

We discuss some point in the case that we adopt $\beta_L$ as one of the invertible base.
We could adopt $\beta_L$ as one of the invertible base since $L(V) \beta_L |_{\lambda =0}$ contains
$T$, not $\dot{T}$.
Although $L(V) \beta_L$ contains $\dot{T}$, the order with respect to the perturbation parameter $\lambda$
of the coefficient of  $\dot{T}$ is larger by one than that of the coefficient of $T$.
Then as long as the gauge vector analytic with respect to the perturbation parameter $\lambda$ is considered,
the choice of $\beta_L$ as one of the base does not spoil the invertible property of the base.

In the main body of this paper, in the zeroth order of the gradient expansion, we discuss the formula 
of the change of the bases of the gauge invariant perturbation variables.
In this appendix, the formula of the change of the bases in the general spacetime is derived.
We consider two arbitrary invertible bases $S_i$ ($i=1,2$)
For $i=1, 2$, let $V_i$ be the gauge vector associated with the invertible base $S_i$.
Clearly the difference of the gauge vectors $V_1 - V_2$ is the background like object.

\paragraph{Theorem}
The difference of the gauge vectors $V_1 - V_2$ can be written in the form of the linear combination 
of $D(S_2) S^a_1$ ($a=1, \cdots ,4$).
In the same way, $V_1 - V_2$ can also be written in the form of the linear combination 
of $D(S_1) S^a_2$ ($a=1, \cdots ,4$).

\paragraph{Proof}
Since 
\begin{equation}
 0 =  D(S_1) S_1 = D(S_2) S_1 - L(V_1-V_2) S_1,
\end{equation}
and since the base $S_1$ is invertible, the difference of the gauge vectors $V_1 - V_2$ can be written 
in the form of the linear combination of $D(S_2) S^a_1$ ($a=1, \cdots ,4$).
In the same way, since 
\begin{equation}
 0 =  D(S_2) S_2 = D(S_1) S_2 + L(V_1-V_2) S_2,
\end{equation}
and since the base $S_2$ is invertible, the difference of the gauge vectors $V_1 - V_2$ can be written 
in the form of the linear combination of $D(S_1) S^a_2$ ($a=1, \cdots ,4$).
We complete the proof.
\vspace{0.5\baselineskip}

For an arbitrary background like object $A$, since 
\begin{equation}
  D(S_1) A = D(S_2) A - L(V_1-V_2) A,
\end{equation}
where the difference of the gauge vectors $V_1 - V_2$ can be written 
in the form of the linear combination of $D(S_2) S^a_1$ ($a=1, \cdots ,4$) from the above theorem,
we can see that the change from the old base $S_1$ to the new base $S_2$ is made possible.
The higher order of the formula of the change of the base, for example, the second order, is given by 
\begin{equation}
 D(S_1)^2 A = D(S_2)^2 A - L(D(S_2) (V_1 - V_2) ) A
- 2 L(V_1-V_2) D(S_2) A + L(V_1-V_2)^2 A.
\end{equation}
In the derivation of the above equation, please notice
\begin{equation}
 D(S_1) [L(W) X] = L(D(S_1) W) X + L(W) D(S_1) X,
\end{equation}
for an arbitrary vector $W$ and an arbitrary tensor $X$.

\section{Proof of Theorem $2.1$}
\label{apdxproof2.1}

The gauge transformation of $D(Z_1, \cdots, Z_n) X$ is given by
\begin{equation}
 \frac{d}{d \mu} D(Z_1, \cdots, Z_n) X 
 = L \left( \frac{d T}{d \lambda} \right) X
 - \sum_{ij} L \left( \frac{d T}{d \lambda} \right) Z_j \cdot
   \frac{d t_i}{d Z_j} \frac{d X}{d t_i}
 + L(T) D(Z_1, \cdots, Z_n) X.
\end{equation}
The first term and the second term in the right hand side cancel since
$X$, $Z_j$ are the scalar like objects.
Therefore $D(Z_1, \cdots, Z_n) X$ is a many time background like object.
Since $(d A / d t_i) / (d B / d t_i)$ is a many time scalar like object 
for many time scalar like objects $A$, $B$ and 
\begin{equation}
  \frac{d t_i}{d Z_j} =  
  \left[ {\rm the \; cofactor \; of \; the \; matrix} 
  \left( \frac{d Z_j}{d t_i} \right) \right] 
  \Bigg/ \det{\left( \frac{d Z_j}{d t_i} \right)},      
\end{equation}
$d t_i / d Z_j \cdot d X / d t_i$ is a many time scalar like object. 
Therefore $D(Z_1, \cdots, Z_n) X$ is a many time scalar like object.

Then Theorem $2.1$ follows.

\section{Representation of the perturbation solution in terms of the many time 
background/scalar like object: concrete example}
\label{apdxconcrete}

In order to have readers understand the representation of the perturbation solutions by the many time 
background/scalar like objects, we give the explanation using the concrete toy model.

In the zeroth order of the gradient expansion, in the universe consisting of the dust fluid, the following
evolution equation holds:
\begin{equation}
 \frac{d \rho}{d t} = - 3 \frac{1}{a}  \frac{d a}{d t} \rho,
\end{equation}
where $\rho$, $a$ are the energy density, the scale factor, respectively.
By letting $t_0$, $t_1$ the initial time, the final time, respectively, the solution to this evolution equation is given by 
\begin{equation}
 F(t_0, t_1) =0, \label{examplebs1}
\end{equation}
where 
\begin{equation}
  F(t_0, t_1) := \rho(t_1) - \rho(t_0) \left( \frac{a(t_0)}{a(t_1)} \right)^3,
\label{examplebs2}
\end{equation}
where all the dependences upon the spatial coordinate ${\bm x}$ are suppressed.
The equations (\ref {examplebs1}),  (\ref {examplebs2}) give the full solution when
$\lambda = 1$.
Since when $\lambda \neq 0$, $F(t_0, t_1)$ is the many time background like object , when  $\lambda = 0$
the equations (\ref {examplebs1}),  (\ref {examplebs2}) give the gauge invariant expression of the background 
level of the solution. 
Since for arbitrary real values $t_0$, $t_1$, the equations (\ref {examplebs1}),  (\ref {examplebs2}) hold,
the following equations hold:
\begin{equation}
  \frac{d}{d t_0} F(t_0, t_1) =  \frac{d}{d t_1}  F(t_0, t_1) =0, 
\end{equation}
whose equations can also be verified by the direct differentiations of the equations  (\ref {examplebs1}),  (\ref {examplebs2}). 
As for an arbitrary set of the many time background/scalar like objects $\{ Y_0, Y_1 \}$ and for an arbitrary natural 
number $k$, we can obtain
\begin{equation}
 D(Y_0, Y_1)^k F(t_0, t_1) = 0,
\end{equation}
which gives the gauge invariant expression of the $k$-th order perturbation of the solution when  $\lambda = 0$, since
its left hand side is the background like object when  $\lambda \neq 0$.
In particular, as the set $\{ Y_0, Y_1 \}$ choosing $S_0$, $S_1$ which are the arbitrary  single time background/scalar like objects,
different in general, at $t_0$ $t_1$, respectively, gives for $k=1$, for example,
\begin{align}
  0 &= D(S_0, S_1) F(t_0, t_1) \notag\\
&= D(S_1) \rho(t_1) - D(S_0) \rho(t_0) \left( \frac{a(t_0)}{a(t_1)} \right)^3 \notag\\
& \quad -3 \rho(t_0) \left( \frac{a(t_0)}{a(t_1)} \right)^3
\left[  \frac{ D(S_0) a(t_0)}{a(t_0)} -
 \frac{ D(S_1) a(t_1)}{a(t_1)}
\right].
\end{align}
This expression gives the relation between $D(S_0) \rho(t_0)$, $D(S_0) a(t_0)$ the gauge invariant perturbations 
corresponding with the perturbations on the time slice $t_0$ where the single time scalar like object  
$S_0 (t_0, {\bm x})$ does not depend on the spatial coordinate ${\bm x}$ and
$D(S_1) \rho(t_1)$, $D(S_1) a(t_1)$ the gauge invariant perturbations 
corresponding with the perturbations on the time slice $t_1$ where the single time scalar like object  
$S_1 (t_1, {\bm x})$ does not depend on the spatial coordinate ${\bm x}$.
This expression corresponds with seeing the physics on the coordinate system whose hypersurfaces defined by
$t=t_i$ ($i=0, 1$) is those on which $S_i (t_i, {\bm x})$ does not depend on the spatial coordinate ${\bm x}$
and whose hypersurfaces  defined by $t$  satisfying $t_0 < t <t_1$ are arbitrary.
I would like the reader to understand that for different times the different arbitrary sets up of the time slices are allowed.

\section{Proof of Theorem $GMW.3$}
\label{apdxproofG6.3}

In the proof we use the Baker Campbell Hausdorff formula given by
\begin{equation}
 \exp{A} \exp{B} =\exp{C},
\end{equation}
where
\begin{equation}
 C = A+B+ \frac{1}{2} [A,B] + \frac{1}{12} [A [A,B]] + \frac{1}{12} [B [B,A]]
 - \frac{1}{24} [A [B [A, B]]]  + \cdots,
\end{equation}
where $C$ is the sum of the terms in which all the products are written by the commutators.
This formula will be referred to as BCH formula from now on.
The equation in the theorem (\ref{equivalencegenerator}) is rewritten as
\begin{equation}
 \exp{[ L(\tilde{V}) ]} \exp{[ \lambda (\frac{d}{d \Lambda} - L(V))]} X \Bigg|_{\Lambda = 0} =
 \exp{[\lambda \frac{d}{d \Lambda}] } X \Bigg|_{\Lambda =0}.
\end{equation}
In the above equation, $\tilde{V}$ is expanded with respect to the perturbation parameter $\lambda$,
on the other hand as for the expansions of $V$, $X$, all the expansion parameters $\lambda$'s are 
replaced with the newly introduced perturbation parameters $\Lambda$'s. 
Using the BCH formula, the above equation can be rewritten as 
\begin{equation}
 \exp{[ \lambda (\frac{d}{d \Lambda} + L(W))]} X \Bigg|_{\Lambda = 0} =
 \exp{[\lambda \frac{d}{d \Lambda}] } X \Bigg|_{\Lambda =0}.
\end{equation}
The vector $W$ depends upon both the expansion parameters $\lambda$, $\Lambda$ and is given by
\begin{equation}
 W = \frac{\tilde{V}}{ \lambda} -V -\frac{1}{2} [\tilde{V}, V] -\frac{1}{12} [\tilde{V} [\tilde{V}, V]]
+ \frac{\lambda}{12} [\tilde{V}, \frac{d V}{d \Lambda}] -\frac{\lambda}{12} [V [\tilde{V}, V]] + \cdots,
\end{equation}
in whose right hand side all the products are written by the commutators.
Furthermore by using the BCH formula the above equality becomes 
\begin{equation}
 \exp{[\lambda \frac{d}{d \Lambda}] }
 \exp{[ \lambda L(W') ]} X \Bigg|_{\Lambda = 0}=
 \exp{[\lambda \frac{d}{d \Lambda}] } X \Bigg|_{\Lambda =0}.
\end{equation}
In order that the above equality holds as for an arbitrary background like object $X$, the equality as
\begin{equation}
 <W'>:= \exp{[\lambda \frac{d}{d \Lambda}]  }W' \Bigg|_{\Lambda = 0}=0
\end{equation}
must hold order by order with respect to the perturbation parameter $\lambda$.
By the above definition, $<W'>$ is given by replacing the new perturbation parameter $\Lambda$ in $W'$
depending upon $\lambda$, $\Lambda$ with the old perturbation parameter $\lambda$.
The BCH formula gives $W'$ as
\begin{equation}
 W' = \sum^{\infty}_{k=0} \frac{(- \lambda)^k}{(k+1)!} \frac{d^k W}{d \Lambda^k}
+ \frac{\lambda^2}{12} [W, \frac{d W}{d \Lambda}]
- \frac{\lambda^3}{24} [W, \frac{d^2 W}{d \Lambda^2}]
+ \cdots,
\end{equation}
where all the products in the right hand side are written by the commutators.
By using the expansions as
\begin{equation}
 \tilde{V} = \sum^{\infty}_{k=1} \frac{\lambda^k}{ k!} \tilde{V}_k, \quad
 V = \sum^{\infty}_{k=0} \frac{\Lambda^k}{k!} V_{k+1},
\end{equation}
and by arranging all the terms not depending upon the commutator by the binomial theorem, 
we obtain $<W'>$ as
\begin{equation}
 <W'> = \sum^{\infty}_{k=0}  \frac{\lambda^k}{ (k+1)!} \tilde{V}_{k+1} -
\sum^{\infty}_{k=0}  \frac{\lambda^k}{ (k+1)!} V_{k+1} + \cdots,
\end{equation}
where $\cdots$ are the sum of the more than two order terms in which all the products are 
written by commutators. 
Then we can prove the equations (\ref{equivalencerelation}). $\Box$

\section{Proof of Theorem $3.1$}
\label{apdxproof3.1}

For $\sigma = \lambda, t_j$, the following equality holds:
\begin{equation}
 \frac{d}{d \sigma} \frac{\partial t_k}{\partial Z_i} =
 - \frac{\partial t_m}{\partial Z_i}
 \frac{d}{d \sigma} \frac{\partial Z_l}{\partial t_m}
 \frac{\partial t_k}{\partial Z_l}.
 \label{usefulformula1}
\end{equation}
This can be obtained by operating $d / d \sigma$ on the equality
\begin{equation}
 \frac{\partial t_k}{\partial Z_j}
 \frac{\partial Z_l}{\partial t_k} =
 \delta_{jl}. 
\end{equation}
For a many time background/scalar like object $X$, we can calculate
\begin{align}
 \left[ D(Y_1, \cdots, Y_n), \frac{\partial}{\partial Z_i} \right] X 
&=\frac{d}{d \lambda} \frac{\partial t_k}{\partial Z_i} \frac{d X}{d t_k}
-\frac{d Y_j}{d \lambda} \frac{\partial t_l}{\partial Y_j}
 \frac{d}{d t_l} \frac{\partial t_k}{\partial Z_i}
 \frac{d X}{d t_k} \notag\\
&+\frac{\partial t_k}{\partial Z_i} 
 \frac{d^2 Y_j}{d t_k d \lambda}  
 \frac{\partial t_l}{\partial Y_j}
 \frac{d X}{d t_l}
+\frac{\partial t_k}{\partial Z_i} 
 \frac{d Y_j}{d \lambda}
 \frac{d^2 t_l}{d t_k d Y_j}  
 \frac{d X}{d t_l}.
\label{leftcommutator} 
\end{align}
On the other hand, we can calculate
\begin{align}
 - \sum_j \frac{\partial}{\partial Z_i}
 [ D(Y_1, \cdots, Y_n)Z_j] \frac{\partial X}{\partial Z_j}
&= - \frac{\partial t_m}{\partial Z_i}
 \frac{d^2 Z_j}{d t_m d \lambda} 
 \frac{\partial X}{\partial Z_j}
+\frac{\partial t_m}{\partial Z_i}
 \frac{d Y_k}{d \lambda}
 \frac{\partial t_l}{\partial Y_k}
 \frac{d^2 Z_j}{d t_m d t_l}
 \frac{\partial X}{\partial Z_j}
\notag\\ 
&+ \frac{\partial t_m}{\partial Z_i}
 \frac{d^2 Y_k}{d t_m d \lambda}
 \frac{\partial t_l}{\partial Y_k}
 \frac{d Z_j}{d t_l}
 \frac{\partial X}{\partial Z_j} 
+ \frac{\partial t_m}{\partial Z_i}
 \frac{d Y_k}{d \lambda}
 \frac{d^2 t_l}{d t_m d Y_k}
 \frac{d Z_j}{d t_l} 
 \frac{\partial X}{\partial Z_j}.
\label{rightcommutator}
\end{align}
By the chain rule and Eq.(\ref{usefulformula1}), for $n=1,\cdots, 4$, we can confirm
that the $n$-th term in the right hand side of Eq.(\ref{leftcommutator}) is equal to
the $n$-th term in the right hand side of Eq.(\ref{rightcommutator}).

Then theorem $3.1$ follows.

\section{Proof of Theorem $3.2$}
\label{apdxproof3.2}

Since we can calculate
\begin{equation}
 D(Y_1, \cdots, Y_n) X = \frac{d X}{d \lambda} - \frac{d Y_j}{d \lambda}
 \frac{d t_i}{d Y_j} \frac{d X}{d t_i},
\end{equation}
\begin{equation}
 D(Z_1, \cdots, Z_n) X = \frac{d X}{d \lambda} - \frac{d Z_j}{d \lambda}
 \frac{d t_i}{d Z_j} \frac{d X}{d t_i},
\end{equation}
and
\begin{equation}
 - \sum_i D(Z_1, \cdots, Z_n) Y_i \cdot \frac{\partial X}{\partial Y_i}
 = - \frac{d Y_i}{d \lambda}
 \frac{d t_l}{d Y_i} \frac{d X}{d t_l}
 + \frac{d Z_j}{d \lambda}
 \frac{d t_k}{d Z_j} \frac{d X}{d t_k},
\end{equation}
the theorem $3.2$ follows.

\section{Proof of Theorem $4.1$}
\label{apdxproof4.1}

We evaluate the difference between the junction model and the DD model;
$\Delta D (N-N(0))$.
The difference of the righthand side of Eq.(\ref{integrationscalefactor})
is given by
\begin{equation}
 \Delta D \frac{1}{3 \rho_3 + 4 \rho_4} \sim \frac{2 r}{R^3} DR
 - \frac{1}{R^2} Dr,
\label{scalefactorerror} 
\end{equation}
where $R$ is $3 \rho_3 + 4 \rho_4$ in the junction model and is of order of
$O(1) \rho$, $r$ is the difference of $3 \rho_3 + 4 \rho_4$ between in the junction
model and in the DD model.
We evaluate the integration of the righthand side of Eq.(\ref{integrationscalefactor}) 
for $\rho_{\ast} \ge \rho$, while for $\rho(1) \ge \rho \ge \rho_{\ast}$ the evaluation 
is much easier since the integration interval is short.
For $\rho_{\ast} \ge \rho$, $R$, $r$ is given by
\begin{align}
 R &= 4 R_{4i} + 3 R_{3r} +4 R_{4r}, \\
 r &= 3 \rho_{3i} + 4 \Delta_{4i} 
\end{align}
where $\Delta_{4i}$ is the difference of $\rho_{4i}$ between in the junction model and 
in the DD model and is given by
\begin{equation}
\Delta_{4i} = - \rho_{3i} (1) \left( \frac{\rho}{\rho(1) }\right) \int^{\infty}_t
 dt \left( \frac{t}{\beta} \right)^{1/2} \exp{(-t + \beta)}.
\end{equation}
$D \rho_{3r}(0)$, $D \rho_{4r} (0)$ are contained by $D R$ and are not by $D r$.
We evaluate the integration by using the inequality as
\begin{equation}
 \Bigg| \int^{\rho_{\ast}}_{0} d \rho \frac{1}{\rho^{l+1}} 
 \left( \frac{\rho}{\rho(1) }\right)^k \exp{(-t + \beta)}         
 \Bigg| \le \left( \frac{\kappa^2}{\Gamma^2_i} \right)^l O(1),
\end{equation}
where $k$, $l$ are finite and $\beta$ is of order of unity.
Along the above outline, Theorem $4.1$ is obtained.

\section{Proof of Eq.(\ref{firstorderbardeen})}
\label{apdxproofrecursive}

As for $\rho(i) \ge \rho \ge \rho(i+1)$, by operating the perturbation operator
$D$ on the defining expressions of the many step reheating, we obtain 
\begin{align}
 &D \Delta N_i = 
\frac{1}{(f_i)_2} D \left( \frac{\Gamma^2_{i+1}}{\Gamma^2_i} \right)
- \frac{(f_i)_1}{(f_i)_2} D s(i),\\
&\frac{D s(i)}{s(i)} = \frac{D \rho_{3i}(0)}{\rho_{3i}(0)}
- \frac{D \Gamma^2_i}{\Gamma^2_i}
- 3 D(N(i)-D(0)).
\end{align}
Please notice
\begin{equation}
 D \Delta N_i = s_i D(N(i)-D(0)) + \cdots.
\end{equation}
We consider how $\alpha$ in $D \Delta N_i$ is transmitted into $D(N-D(0))$.
Putting $p_i = D \Delta N_i$, we obtain
\begin{align}
 p_i &= \alpha,\notag\\
 p_{i+1} &= s_{i+1} p_i, \notag\\
 p_{i+2} &= s_{i+2} (p_i + p_{i+1}), \notag\\
 &\cdots \notag\\
 p_{M} &= s_{M} (p_i + p_{i+1}+ \cdots + p_{M-1} ).
\end{align}
From them, we obtain the recursive relation for $q_k = \sum^k_{j=i} p_j$ as
\begin{align}
 &q_{k+1} = (s_{k+1} +1) q_k \quad k=i, i+1, \cdots, M-1,\notag\\
 &q_i = p_i = \alpha.
\end{align}
It can be easily solved as
\begin{equation}
 q_k = (s_k + 1)(s_{k-1} + 1) \cdots (s_{i+1} +1) \alpha
\end{equation}
We adopt $ - s_i / 3 \cdot D \rho_{3i}(0) / \rho_{3i}(0)$ as $\alpha$.

Since $D \Gamma^2_{i+1} / \Gamma^2_{i+1}$ appears in $p_i$, $p_{i+1}$, 
we obtain
\begin{align}
 p_i &= \alpha,\notag\\
 p_{i+1} &= \beta + s_{i+1} p_i, \notag\\
 p_{i+2} &= s_{i+2} (p_i + p_{i+1}), \notag\\
 &\cdots \notag\\
 p_{M} &= s_{M} (p_i + p_{i+1}+ \cdots + p_{M-1} )
\end{align}
In the same way as the above, it can be solved as
\begin{equation}
 q_k = (s_k + 1)(s_{k-1} + 1) \cdots (s_{i+2} +1) [(s_{i+1} +1) \alpha + \beta].
\end{equation}

\section{Exponent evaluation}
\label{apdxexponent}

In this appendix, for notational simplicity, we will simply write that $A \sim B \sim 10^{- \alpha}$ 
omitting the absolute value marks $|\cdots|$,
when we intend to imply $|A| \sim |B| \sim 10^{- \alpha}$ where $A$, $B$ are some physical quantities 
and $\alpha$ is the positive number.
Since based on this notation we can evaluate
\begin{align}
 &(f_i) \sim (f_i)_2 \sim (g_i) \sim (g_i)_2 \sim 10^{- k_i} 10^{-3 n_i} + 10^{- 4 n_i}
 \sim 10^{- 3 n_i - m_i},\notag\\
 &(f_i)_1 \sim (g_i)_1 \sim 10^{- 3 n_i}, \quad {\rm etc},
\end{align}
we can obtain
\begin{align}
 &s_i  = 3 s(i) \cdot \frac{ \exp{[- 3 \Delta N_i]}- \exp{[- 4 \Delta N_i]}}
   {- 3 s(i)\exp{[- 3 \Delta N_i]} - 4 (1-s(i))\exp{[- 4 \Delta N_i]}}
\sim 10^{- k_i + m_i},\\
 &1+s_i = 
 \frac{(s(i)-4) \exp{[- 4 \Delta N_i]}}
   {- 3 s(i)\exp{[- 3 \Delta N_i]} - 4 (1-s(i))\exp{[- 4 \Delta N_i]}}
 \sim 10^{- n_i + m_i},\\
 &F_{i} \sim 10^{- k_{i}+m_{i}} (\delta_{i}-1)
 \sim (\delta_{i}-1) s_{i}.
\end{align}
The derivatives of $s_i$ with respect to the variable $1$ ($s(i)$), 
the variable $2$ ($\Delta N_i$) are evaluated as
\begin{align}
 &(s_i)_1 s(i) = \frac{3 s(i)}{(f_i)^2_2} (f_i)_1 (-4) \exp{[- 4 \Delta N_i]}  
  \sim 10^{- n_i - k_i+2 m_i} \sim (s_i) (1+s_i),\\  
 &(s_i)_2 = \frac{3 s(i)}{(f_i)^2_2} (s(i)-4) \exp{[- 7 \Delta N_i]}
 \sim 10^{- n_i - k_i+2 m_i} \sim (s_i) (1+s_i).
\end{align}
Next we consider the derivatives of $F_i$ with respect to the variable $1$ ($s(i)$), 
the variable $2$ ($\Delta N_i$) and the variable $\delta_i$.
The derivatives of $F_i$ with respect to the variable $1$, the variable $2$ 
are calculated as
\begin{align}
 &(F_i)_1 s(i) = (\delta_{i}-1) \exp{[- 3 \Delta N_i]} s(i)
 \frac{A_i}{(f_i)^2_2 (g_i)^2},\\ 
 &(F_i)_2 = (\delta_{i}-1) \exp{[- 3 \Delta N_i]} s(i) 
 \frac{B_i}{(f_i)^2_2 (g_i)^2}, 
\end{align}
where
\begin{align}
 &A_i = \{ (f_i)_1 s(i) + (f_i) \} (f_i)_2 (g_i)
- (f_i) \{ s(i) (f_i)_{12} (g_i) - s(i) (f_i)_2 (g_i)_1 \},\\  
 &B_i = \{ (f_i)_2 - 3 (f_i) \} (f_i)_2 (g_i)
- (f_i) \{ (f_i)_{22} (g_i) + (f_i)_2 (g_i)_2 \}.
\end{align}
Putting $a_i = \exp{(-3 \Delta N_i)}$, $b_i = \exp{(-4 \Delta N_i)}$
$A_i$, $B_i$ can be expanded as
\begin{equation}
 A_i, B_i \sim \left( s(i) a_i \right)^3
+ \left( s(i) a_i \right)^2 b_i
+ \left( s(i) a_i \right) b^2_i
+ b^3_i,
\end{equation}
where the coefficients are of order of unity.
Since we can verify that the terms of $\left( s(i) a_i \right)^3$ are vanishing
by the manifest calculations, we can evaluate 
\begin{equation}
 A_i, B_i \sim  \left( s(i) a_i \right)^2 b_i
+ \left( s(i) a_i \right) b^2_i
+ b^3_i \sim 10^{- 10 n_i - 2 m_i},
\end{equation}
which give the estimations as 
\begin{equation}
 (F_i)_1 s(i) \sim (F_i)_2 \sim (\delta_i -1) 10^{- k_i+m_i} 10^{-n_i+m_i}
 \sim (\delta_i -1) s_i (1+s_i).
\end{equation}
On the other hand, the derivative of $F_i$ with respect to the variable $\delta_i$
is estimated as
\begin{equation}
 \frac{d}{d \delta_i} F_i = \frac{(f_i) s(i) \exp{[-3 \Delta N_i]}}{(g_i)^2}
 \sim 10^{- k_i +m_i} \sim s_i.
\end{equation}

\section{Property of oblivion}
\label{apdxoblivion}

First we will verify that in the multiple fluid system without the metric junction
the property of oblivion holds.
We consider the situation in which the energy of one component $\rho_1$ is
dominant in a many components system. 
We will show that the Bardeen parameter, that is
the perturbation of $\Delta N = N - N(0)$ is governed by the perturbation of
the energetically dominant component $\rho_1$ and the contribution to it from
the energetically subdominant component $\rho_2$ is suppressed by 
$\rho_2 / \rho_1$.
Please notice that $\rho_2 / \rho_1$ is the energy ratio at the present time, 
not that at some past time.
Even if $\rho_2$ governed the cosmic energy in the past, the contribution to
the Bardeen parameter from $\rho_2$ at the present time is small. 
Since the Bardeen parameter forgets the energy history of the universe that 
$\rho_2$ governed the cosmic energy in the past, we call this property the property 
of oblivion.

Such property is possessed by the multiple fluids system without the metric junction.
We assume that the $k$ fluid governs the cosmic energy in the system in which the
$k$ fluid and the $l$ fluid coexists.
By simple calculation, the perturbative calculation with respect to $\rho_l (0)$
of $\Delta N$ is given by
\begin{equation}
 \Delta N = - \frac{1}{k} \ln{\frac{\rho}{\rho_k (0)}} + \Delta N_1 + \Delta N_2
 + \cdots,
\end{equation}
where $\Delta N_i$ is the correction term proportional to $\rho_l (0)^i$ and is given by
\begin{align}
 &\Delta N_1 = \frac{\partial \Delta}{\partial \rho_l (0)} \Bigg|_0 \rho_l (0)
= \frac{1}{k} \frac{\exp{[-l \Delta N]}}{\rho_k (0) \exp{[-k \Delta N]}} \rho_l (0)
\sim \frac{\rho_l}{\rho_k}    ,\\
 &\Delta N_2 =\frac{1}{2}\frac{\partial^2 \Delta}{\partial \rho_l (0)^2} \Bigg|_0 
 \rho_l (0)^2 =\frac{1}{2} \left( \frac{1}{k} - 2 \frac{l}{k^2} \right)
 \frac{\exp{[-2 l \Delta N]}}{\rho_k (0)^2 \exp{[-2 k \Delta N]}} \rho_l (0)^2 
\sim \frac{\rho^2_l}{\rho^2_k},
\end{align}
where $|_0$ implies the differentiation at $\rho_l (0) =0$.
This property of oblivion follows from the fact that each component satisfies
\begin{equation}
 \frac{d \rho_i}{d N} = c_i \rho_i,
\end{equation}
where $c_i$ is a constant of order of unity.

The property of oblivion is not possessed by the system composed by the multiple
slow rolling scalar fields with quadratic potentials.
We assume that $\phi_1$, $\phi_2$ are dominant, subdominant, respectively. 
$\Delta \tau = \tau - \tau (0)$ is given by
\begin{equation}
 \Delta \tau = - \frac{1}{2 m^2_1} \ln{\frac{\rho}{\rho_1 (0)}} 
+ \Delta \tau_1 + \Delta \tau_2 + \cdots,
\end{equation}
where $m_a$ is the mass of $\phi_a$,
$\Delta \tau_i$ is the correction term proportional to $\phi_2 (0)^i$ and 
is evaluated as
\begin{equation}
 \Delta \tau_1 \sim \frac{1}{m^2_1} \frac{\rho_2}{\rho_1}, \quad
 \Delta \tau_2 \sim \frac{1}{m^2_1} \left( 1+ \frac{m^2_2}{m^2_1} \right)
 \frac{\rho^2_2}{\rho^2_1},
\end{equation}
where all the coefficients of order of unity are suppressed.
Since $\Delta N$ is a function of $\Delta \tau$, $\Delta N$ does not have
the property of oblivion.
In fact, the correction of $\Delta N$ proportional to $\rho_2 (0)$ contains
the terms of order of $m^2_2 / m^2_1 \cdot \rho_2 / \rho_1$.
Then when $\rho_2 / \rho_1 \ll 1$ but $m^2_2 / m^2_1 \gg 1$, the term originating
from $\rho_2$ is memorized.
That this system does not have the property of oblivion follows from the fact that
each component experiences the dilution depending on $m^2_a$ as 
\begin{equation}
 \frac{d \rho_a}{d \tau} = - 2 m^2_a \rho_a. 
\end{equation}

It is not self-evident if the property of oblivion holds in the system with 
the metric junction.
As the example where the property of oblivion does not hold, we can give 
the system which has the slow rolling phase in the first period and the fluid phase 
in the second period connected to the first period by the metric junction.
As the toy model, we consider the following situation;
In the first period $t_0 < t <t_1$, the scalar fields $\phi_1$, $\phi_2$ with the 
equal masses $m$ are in the slow rolling phase.
On the surface defined by $\rho(1) = c$ where $c$ is the constant which can be 
ordinarily written in terms of the masses or the decay constants of the scalar fields,
$\phi_1$ is transformed into the $3$ fluid $\rho_1$ and 
$\phi_2$ is transformed into the $4$ fluid $\rho_2$.
In the slow rolling phase, the energy of $\rho_2$ is dominant, but at the present time 
in the fluid phase, the energy of $\rho_1$ is dominant, that is 
\begin{align}
 &\frac{\rho_1 (0)}{\rho_2 (0)} \ll 1,\\
 &\frac{\rho_2}{\rho_1} = \frac{\rho_2 (0)}{\rho_1 (0)} e^{- \Delta N} \ll 1,
\end{align}
where $\Delta N = N - N(1)$.
Putting  $D = D(\rho, \rho(1), a(0))$,
the Bardeen parameter at the present time $t$ is given by
\begin{equation}
 D(\rho) N = D (N(1)-N(0)) + D(N - N(1)),
\end{equation}
where
\begin{equation}
 D (N(1)-N(0)) = \frac{\kappa^2}{2 m^2} D \rho_1 (0)
 + \frac{\kappa^2}{2 m^2} D \rho_2 (0), \label{firststep}
\end{equation}
and
\begin{align}
 D (N(1)-N(0)) &= - \frac{D \rho (0)}{\rho (0)}
 \frac{\rho_1 (0) e^{-3 \Delta N} + \rho_2 (0) e^{-4 \Delta N}}
{3 \rho_1 (0) e^{-3 \Delta N} + 4 \rho_2 (0) e^{-4 \Delta N}}
\notag\\
& +
\frac{D \rho_1 (0) e^{-3 \Delta N} + D \rho_2 (0) e^{-4 \Delta N}}
{3 \rho_1 (0) e^{-3 \Delta N} + 4 \rho_2 (0) e^{-4 \Delta N}}
\notag\\ 
 &\cong - \frac{1}{3} \frac{D \rho_2 (0)}{\rho_2 (0)}
 + \frac{1}{3} \frac{D \rho_1 (0)}{\rho_1 (0)},
 \label{secondstep}
\end{align}
where
\begin{equation}
 \rho_a (0) = \frac{1}{2} m^2 \phi_a^2 (0), \quad (a=1,2).
\end{equation}
We can find that the effect of the perturbation of the component energetically dominant in the past
$D \rho_2 (0)$ in (\ref{firststep}) is not canceled by the effects of the perturbations in the 
fluid phase (\ref{secondstep}), although $\rho_2$ has already become energetically 
subdominant, that is, such perturbation remains in the memory and that the property of 
oblivion breaks.

\section{the property of oblivion of the many step reheating 
in the viewpoint of the change of the base and relation with other approaches}

\label{apdxchangebase}

Received by the request of the reviewer, we will discuss the relation between our formulation
and the $\delta N$ formalism \cite{Sasaki2006}, the separate universe approach \cite{Wands2000}.
Since these papers \cite{Sasaki2006} \cite{Wands2000} mainly treats many fluid system, in this appendix 
we will also treat such system.
In order to derive the results which can be compared with those of the paper \cite{Sasaki2006},
we have judged that it is the most appropriate to use the formula of the change of the base, theorem 
\ref{mainchangebase}.$2$ formulated in the section \ref{mainchangebase}.
Therefore it is necessary to demonstrate how effective the formula of the change of the base is
in the concrete example; many step reheating.
While in the section \ref{mainmanystep} we determined the time evolution of the Bardeen parameters
by solving repeatedly the recursive relations between one energy transfer and the next energy transfer,
in this appendix we will solve the many step reheating by using the change of the base 
theorem \ref{mainchangebase}.$2$. 
Next we derive the main results of the paper \cite{Sasaki2006} by using the change of the base 
theorem \ref{mainchangebase}.$2$. 
At last, we reinterpret the separate universe approach \cite{Wands2000} using our notation and 
rederive the main results of the paper \cite{Wands2000} by our framework.

In this appendix, although we use the concept of the many time background/scalar like objects,
as their base we will not use any many time scalar like objects belonging to the plural different times such as
$S(t_1, t_2)= \ln{(a(2)/a(1))}$.
We consider the base $\{S_k (t_k)\}^n_{k=1}$ where $S_k (t_k)$ is the single time scalar like object 
at the single time $t_k$. 
For an arbitrary many time background/scalar like object $A(t_1, t_2, \cdots, t_n)$, the new many time
background/scalar like object is constructed as
\begin{equation}
 D(S_1(1), S_2(2),\cdots, S_n(n)) A := 
\left[ \frac{d}{d \lambda} - \sum^n_{k=1}  
\left( \frac{d S_k (t_k)}{d \lambda} \Big/ \frac{d S_k (t_k)}{d t_k}\right)  \frac{d}{d t_k} \right] A.
\end{equation}
The solution is always written as $A(t_1, t_2, \cdots, t_n) =0$ where $A$ is a function of the single time
scalar like objects $\{ W_k (t_k) \}$.
For an arbitrary single time background/scalar like object $W_k (t_k)$, $W_k (t_k)$
with $D(S_1(1), S_2(2),\cdots, S_n(n))$ operated on is reduced to
\begin{align}
& \quad D(S_1(1), S_2(2),\cdots, S_n(n)) W_k (k)  \notag\\
& = \left[ \frac{d}{d \lambda} - 
\left( \frac{d S_k (t_k)}{d \lambda} \Big/ \frac{d S_k (t_k)}{d t_k}\right)  \frac{d}{d t_k} \right] W_k (t_k)
=: D(S_k (k)) W_k.
\end{align}

We discuss the many step reheating treated in the section  in the viewpoint 
of the change of the base formulated in the section \ref{mainchangebase}.
We consider the sequence of the times $\{t_i\}^M_{i=0}$ satisfying $t_i < t_j$ when $0 \le i < j \le M$.
When $t_0 < t < t_1$, all the fluid components are assumed to be the $3$ fluids  $\{\rho_{3i}\}^M_{i=0}$.
At $t=t_i$, the $3$ fluid $\rho_{3i}$ is converted into the $4$ fluid $\rho_{4i}$ 
with the decay constant $\Gamma^2_i$.
This parametrization $\rho_{3i}(0)$ is a little different from that of the section \ref{mainmanystep},
although this appendix treats the same problem as the section \ref{mainmanystep}.
The decay constants $\Gamma^2_i$ ($i=1,2, \cdots M$) are assumed to fluctuate 
since they are the functions of the modulating fields.
Then when $t_M < t$, all the fluid components are the $4$ fluids  $\{\rho_{4i}\}^M_{i=0}$.

When $t_i < t < t_{i+1}$, since the energy density $\rho$ is given by
\begin{align}
 \rho &= \sum_{j \ge i+1} \rho_{3j} + \sum_{j \le i} \rho_{4j} \notag\\
 &= \sum_{j \ge i+1} \rho_{3j} (0) \left( \frac{a(0)}{a} \right)^3 
 + \sum_{j \le i} \rho_{3j}(0) \frac{a(0)^3 a(j)}{a^4},
\end{align}
using the formula of the change of the base, that is, theorem \ref{mainchangebase}.$2$ 
in the section \ref{mainchangebase}, the Bardeen parameter can be written in the following form:
\begin{equation}
 D(\rho) \ln{a} = D(a) \rho \Big/ \left( - a \frac{d}{d a}\rho \right),
\end{equation}
where the numerator and the denominator in the right hand side are given by
\begin{align}
D(a) \rho &= 
 \sum_{j \ge i+1} D(a) (\rho_{3j} (0)  a(0)^3 ) \frac{1}{a^3}  
 + \sum_{j \le i} D(a)(\rho_{3j}(0) a(0)^3 a(j)) \frac{1}{a^4},\notag\\
- a \frac{d}{d a}\rho &=
3 \sum_{j \ge i+1} \rho_{3j} (0) \left( \frac{a(0)}{a} \right)^3 
 + 4 \sum_{j \le i} \rho_{3j}(0) \frac{a(0)^3 a(j)}{a^4},
\end{align}
where $D(\rho):=D(S(0), C_1, \cdots, C_i, \rho)$ and $D(a):=D(S(0), C_1, \cdots, C_i, a)$.
$S(0)$ is the scalar like object at $t=t_0$ and $C_j$ characterizing the space like surface
where the $3$ fluid $\rho_{3j}$ is converted into the $4$ fluid, is given by
\begin{equation}
 C_j := \frac{\kappa^2}{3} \rho(j) - \Gamma^2_j,
\end{equation}
where $\rho(j)$ is the energy density at $t=t_j$.
Assuming $\delta$ to be the sufficiently small quantity, we put the following assumptions:
\begin{align}
 | D(S(0)) \ln{a(0)} | &\sim \delta,\notag\\
 | D(S(0)) \ln{\rho_{3i} (0)} | &\le \delta,\notag\\
 | D(C_i) \ln{\Gamma^2_i} | &\le \delta,
\end{align}
where $D(C_i):=D(S(0), C_1, \cdots, C_i)$.
Using the formula of the change of the base, theorem \ref{mainchangebase}.$2$ 
in the section \ref{mainchangebase} gives
\begin{equation}
 D(C_i) \ln{a(i)} = D(a(i)) C_i \Big/ \left( - a(i) \frac{d}{d a(i)} C_i \right),
\label{junctionmoto}
\end{equation}
where the new base is $D(a(i)):=D(S(0), C_1, \cdots, C_{i-1}, a(i))$,
whose right hand side is interpreted to be the limit $t \to t_i - 0$ and therefore whose numerator and 
denominator in the right hand side are given by
\begin{align}
D(a(i)) C_i &= 
 \sum_{k \ge i} D(a(i)) (\rho_{3k} (0)  a(0)^3 ) \frac{1}{a(i)^3}  
 + \sum_{k \le i-1} D(a(i))(\rho_{3k}(0) a(0)^3 a(k)) \frac{1}{a(i)^4}\notag\\
 &\quad - \frac{3}{\kappa^2} D(a(i)) \Gamma^2_{i}
,\notag\\
- a(i) \frac{d}{d a(i)} C_i &=
3 \sum_{k \ge i} \rho_{3k} (0) \left( \frac{a(0)}{a} \right)^3 
 + 4 \sum_{k \le i-1} \rho_{3k}(0) \frac{a(0)^3 a(k)}{a(i)^4}.
\label{junctionsaki}
\end{align}
By using the equations (\ref {junctionmoto}), (\ref {junctionsaki}), the induction with respect to 
the natural number $i$ gives 
\begin{equation}
  | D(C_i) \ln{a(i)} | \le \delta,
\end{equation}
for an arbitrary natural number $i$.

For $t_i < t <t_{i+1}$, assuming $\rho_{3j}$ to be the energetically dominant component,
the Bardeen parameters are given by 
\begin{equation}
 D(\rho)^k \ln{a} = \frac{1}{3} D(S(0))^k \ln{\rho_{3j} (0)} + D(S(0))^k \ln{a(0)}.
\end{equation}
On the other hand, assuming $\rho_{4j}$ to be the energetically dominant component, 
the Bardeen parameters are given by 
\begin{equation}
 D(\rho)^k \ln{a} = \frac{1}{4} D(S(0))^k \ln{\rho_{3j} (0)} + \frac{3}{4} D(S(0))^k \ln{a(0)}
 + \frac{1}{4} D(C_j)^k \ln{a (j)}.
\end{equation}
As for the third term in the right hand side of the above equation, 
using the equations (\ref {junctionmoto}), (\ref {junctionsaki})  gives
\begin{align}
 D(C_j) \ln{a(j)} &= \frac{1}{3} D(S(0)) \ln{\rho_{3j} (0)} + D(S(0)) \ln{a(0)} \notag\\
 &- \left( D(C_j) \Gamma^2_j \right) \Bigg/ 
    \left( \kappa^2 \rho_{3j} (0) \left( \frac{a(0)}{a(j)} \right)^3 \right)
\end{align}
since the $j$-th component $\rho_{3j}$, $\rho_{4j}$ is also energetically dominant 
in the earlier periods from $t \to t_j - 0$ to $t = t_i$,
if $\rho_{4j}$ is energetically dominant for  $t_i < t <t_{i+1}$.
By using the above equation recursively, as the higher order perturbation, we obtain
\begin{align}
 D(C_j)^2 \ln{a(j)} &= \frac{1}{3} D(S(0))^2 \ln{\rho_{3j} (0)} + D(S(0))^2 \ln{a(0)} \notag\\
 & \quad
- \left( D(C_j)^2 \Gamma^2_j \right) \Bigg/ 
    \left( \kappa^2 \rho_{3j} (0) \left( \frac{a(0)}{a(j)} \right)^3 \right) \notag\\
& \quad + 3 \left[ \left( D(C_j) \Gamma^2_j \right) \Bigg/ 
    \left( \kappa^2 \rho_{3j} (0) \left( \frac{a(0)}{a(j)} \right)^3 \right) \right]^2.
\end{align}
In the same way as the above, we can also calculate $D(C_j)^k \ln{a(j)}$ ($k=3,4,\cdots$).

We can see that the Bardeen parameters $D(\rho)^k \ln{a}$ are governed by the fluctuation
of the energetically dominant component at the time.
Therefore if the energetically dominant component in the earlier period becomes the energetically 
subdominant at the present time, the fluctuation of such component is forgotten in  
the Bardeen parameters $D(\rho)^k \ln{a}$.
We call this property the property of oblivion in the many step reheating.

Next we discuss the relation with the papers \cite{Sasaki2006} \cite{Wands2000}
written under the name of the $\delta N$ formalism, the separate universe approach, respectively.
From now on, we does not consider the metric junction.

In order to obtain the results which can be compared with the results of the paper \cite{Sasaki2006},
we will set the same situation as in this paper.
We consider the universe where the pressureless dust fluid $\rho_3$ and the radiation $\rho_4$ coexist.
We assume that $S(0)$ is an arbitrary scalar like object at the initial time $t = t_0$.
For the natural number $m=1,2, \cdots$,
the perturbation variables with the scale factor $a$ as the base
\begin{equation}
 D(a)^m \ln{\rho_k} = D(S(0))^m \ln{[ \rho_k (0) a(0)^k ]},
\end{equation}
where $k=3,4$ and the perturbation variables with the energy density of the $l$ fluid 
$\rho_l$ ($l=3$ or $4$) as the base
\begin{align}
 D(\rho_l)^m \ln{a} &= \frac{1}{l} D(S(0))^m \ln{[ \rho_l (0) a(0)^l ]},\notag\\
 D(\rho_l)^m \ln{\rho_k} &= \frac{1}{k} D(S(0))^m \ln{\left[ \frac{\rho_k (0) ^k}{\rho_l (0)^l} \right]},
\end{align}
where $k=3,4$, are the conserved quantities since they can be written in terms of the perturbation  
variables at the initial time only.
Please notice $D(a):= D(a, S(0))$, $D(\rho_l) :=D(\rho_l, S(0)) $.
In order to solve the time evolution of the Bardeen parameters $D(\rho)^m \ln{a}$, every time
we operate $D(\rho)$ on the logarithm of the scale factor $\ln{a}$, we perform the change of 
the base $\rho \to a$, $\rho \to \rho_l$
($l=3$ or $4$), respectively, by applying the theorem \ref{mainchangebase}.$2$ presented in the section 
\ref{mainchangebase}.
By this process, we can write the Bardeen parameters $D(\rho)^m \ln{a}$
as the polynomial of $D(a)^p \ln{\rho_k}$ ($k=3,4$ ; $p=1,2,\cdots, m$),
as the polynomial of $D(\rho_l)^p \ln{a}$, $D(\rho_l)^q \ln{\rho_k} $ ($k=3,4$; $p,q=1,2,\cdots, m$), respectively.
All the coefficients are polynomials of $r$ defined by
\begin{equation}
 r := \frac{3 \rho_3}{3 \rho_3 + 4 \rho_4}.
\label{rationdefinition}
\end{equation}
So far we considered changing the base of the Bardeen parameters $D(\rho)^m \ln{a}$, that is the energy
density of the total system $\rho$ into another single base such as the scale factor $a$, the energy density
of some constituent component $\rho_l$ ($l=3$ or $4$), respectively.
On the other hand, the paper \cite{Sasaki2006} unifies the true value of the perturbation variables 
into the logarithm of the scale factor $\ln{a}$.
In the $\delta N$ formalism \cite{Sasaki2006}, the authors give relations among the perturbations of  
$N:=\ln{a}$ with different bases. 
The method adopted by the paper \cite{Sasaki2006} will be rewritten in our notation.
By using the equation as 
\begin{align}
 D(\rho) \ln {\rho_k} &= D(S(0)) \ln{[\rho_k (0) a(0)^k]} - k D(\rho) \ln{a} \notag\\
&= k D(\rho_k) \ln{a} - k D(\rho) \ln{a}, 
\end{align}
for $k=3,4$ and using the fact that $D(\rho_k) \ln{a}$ is the conserved quantity written in terms of 
the perturbation variables at the initial time $t=t_0$, we obtain
\begin{equation}
 D(\rho) p_{km} = p_{k (m+1)}, \quad 
 D(\rho) \rho_k = \rho_k \> k \> p_{k1},
\label{sasakirecursive}
\end{equation}
where $p_{km}$ is defined by
\begin{equation}
 p_{km} := D(\rho_k)^m \ln{a} -  D(\rho)^m \ln{a}.
\end{equation}
By applying $D(\rho)$ to $\rho_k$ repeatedly with the use of the equations (\ref{sasakirecursive}),
we obtain 
\begin{align}
 D(\rho)^2 \rho_k &= \rho_k [ k^2 p^2_{k1} +k p_{k2} ], \notag\\
 D(\rho)^3 \rho_k &= \rho_k [ k^3 p^3_{k1} +3 k^2 p_{k1} p_{k2} + k p_{k3}], \notag\\
 D(\rho)^4 \rho_k &= \rho_k [ k^4 p^4_{k1} +6 k^3 p^2_{k1} p_{k2} + 4 k^2 p_{k1} p_{k3} + 3 k^2 p^2_{k2}
 + k p_{k4}].
\end{align}
By applying $D(\rho)$ to $\rho_3 + \rho_4 = \rho$ repeatedly with the use of the above equations, 
we can write the Bardeen parameters $D(\rho)^m \ln{a}$ 
as the polynomial of $D(\rho_k)^p \ln{a}$ ($k=3,4$ ; $p=1,2,\cdots, m$).
In the same way as in the above cases, all the coefficients are polynomials  of $r$ defined 
by (\ref{rationdefinition}).

Another derivation of the second equation of (\ref{sasakirecursive}) will be explained.
For $k =3,4$, the following equality holds:
\begin{equation}
 \ln{\rho_k} + k \ln{a} = \ln{[\rho_k (0) a(0)^k]}.
\end{equation}
The right hand side in the above equation is the constant not depending on the present time $t$.
Then the left hand side with $D(\rho)$, $D(\rho_k)$, $D(a)$ operated on is all equal:
\begin{equation}
 D(\rho) \ln{\rho_k} + k D(\rho) \ln{a}  = k D(\rho_k) \ln{a} = D(a) \ln{\rho_k}    
 = D(S(0)) \ln{[\rho_k (0) a(0)^k]}
\end{equation}
From this, we can obtain the second equation of (\ref{sasakirecursive}).
This method is the strong method which gives the relations among the perturbations of the 
same true value with different bases.

By the change of the base $\rho \to \rho_3$ as explained in the above, we will reproduce the 
results of the paper \cite{Sasaki2006}.
We set the same initial condition of the perturbations as in the paper  \cite{Sasaki2006} where 
only the energy density of the pressureless fluid perturbs in the initial time $t = t_0$.
Assuming that $S(0)$ is the scalar like object at the initial time $t = t_0$, the initial perturbations 
are given by
\begin{equation}
 D(S(0))^k \rho_4 (0) = D(S(0))^k a(0) =0, \quad (k=1,2,\cdots),
\end{equation}
and 
\begin{equation}
 D(S(0)) \chi (0) \neq 0, \quad D(S(0))^l \chi(0) =0, \quad (l=2, 3,\cdots),
\end{equation}
as for the $3$ fluid as
\begin{equation}
 \rho_3 (0) = \frac{1}{2} m^2 \chi(0)^2,
\end{equation}
where $m$ is the constant representing the curvaton mass and $\chi(0)$ is the initial value
of the curvaton oscillation.
In this case,  the following equations holds:
\begin{align}
 D(\rho_3) \ln{a} &= \frac{2}{3} \frac{1}{\chi(0)} D(S(0)) \chi (0) =: \zeta_3, \notag\\
 D(\rho_3)^2 \ln{a} &= - \frac{3}{2} \zeta^2_3, \notag\\
 D(\rho_3)^3 \ln{a} &= \frac{9}{2} \zeta^3_3,
\end{align}
where $D(\rho_3):= D(\rho_3, S(0))$ and we notice the constancy of $\zeta_3$.
The formula of the change of the base, that is, theorem \ref{mainchangebase}.$2$ 
in the section \ref{mainchangebase} gives the first order 
Bardeen parameter as
\begin{equation}
 D(\rho) \ln{a} = D(\rho_3) \ln{a} - D(\rho_3) \rho \cdot \frac{d \ln{a}}{d \rho} = r \zeta_3,
\label{mycompare1}
\end{equation}
where $D(\rho):= D(\rho, S(0))$ and (\ref{rationdefinition}).
In the derivation of the above equation, please notice
\begin{equation}
  D(\rho_3) \rho_4 = - 4 \rho_4 \zeta_3.
\end{equation}
By using the equation given by the formula of the change of the base 
in the section \ref{mainchangebase} 
\begin{equation}
 D(\rho) \rho_3 = D(\rho_3) \rho_3 - D(\rho_3) \rho \cdot \frac{d \rho_3}{d \rho} = 4 \rho_4 r \zeta_3,
\end{equation}
we obtain
\begin{equation}
 D(\rho) r = (3+r) (1-r) r \zeta_3.
\label{rderivative}
\end{equation}
By operating $D(\rho)$ on (\ref{mycompare1}) repeatedly using (\ref{rderivative}) and 
the relation as
\begin{equation}
 D(\rho) \zeta_3 = - \frac{3}{2} \zeta^2_3,
\end{equation}
we obtain

\begin{align}
D(\rho)^2 \ln{a}  &= \left[ \frac{3}{2 r} -2 -r \right] r^2 \zeta^2_3, \label{mycompare2}\\
D(\rho)^3 \ln{a}  &= \left[ - \frac{9}{r}  + \frac{1}{2}+ 10 r +3 r^2 \right] r^3 \zeta^3_3.\label{mycompare3} 
\end{align}
The expressions of the first, second, the third orders of the Bardeen parameters
 (\ref{mycompare1}) (\ref{mycompare2}) (\ref{mycompare3}) which we could obtain completely agree 
with the results of the paper \cite{Sasaki2006}.

We extend the above discussion to the more general case that at the initial time $t=t_0$
the energy density of the $4$ fluid $\rho_4 (0)$, the scale factor $a(0)$ as well as the energy density of the 
$3$ fluid $\rho_3 (0)$ fluctuate.
We consider the change of the base $S(0), \rho \to S(0), a$ where $S(0)$ is the arbitrary scalar like object
at the initial time.
We obtain
\begin{equation}
 D(\rho) \rho_3 = - D(\rho) \rho_4 = r_4 D(a) \rho_3 - r_3 D(a) \rho_4,
\end{equation}
where
\begin{equation}
 r_3 := \frac{3 \rho_3}{3 \rho_3 + 4 \rho_4}, \quad
 r_4 := \frac{4 \rho_4}{3 \rho_3 + 4 \rho_4}.
\end{equation}
In this paragraph, we simply write $D(S(0), \rho)$, $D(S(0), a)$ by $D(\rho)$, $D(a)$, respectively. 
Taking into account
\begin{equation}
 D(a) \rho_i = \zeta_{i,1} \rho_i, \quad (i=3,4), 
\end{equation}
where
\begin{equation}
  \zeta_{i,k} := D(S(0))^k \ln{[\rho_i(0) a(0)^i]}, \quad (i=3,4; \; k=1,2,\cdots),
\end{equation}
we can derive
\begin{equation}
 D(\rho) r_3 = - D(\rho) r_4 = 12 \left( \frac{1}{3} r_3 + \frac{1}{4} r_4 \right)
 r_3 r_4 \left( \frac{1}{3} \zeta_{3,1} - \frac{1}{4} \zeta_{4,1} \right).
\label{rparameterderivative}
\end{equation}
By using the formula of the change of the base $\rho \to a$, we obtain the first order Bardeen 
parameter and 
by operating $D(\rho)$ to the first order Bardeen parameter $D(\rho) \ln{a}$ with use of 
(\ref {rparameterderivative}), we obtain the second order Bardeen parameter:
\begin{align}
 D(\rho) \ln{a} &= \frac{1}{3} r_3 \zeta_{3,1} + \frac{1}{4} r_4 \zeta_{4,1},
\label{firstsecondfour}\\
 D(\rho)^2 \ln{a} &= \frac{1}{3} r_3 \zeta_{3,2} + \frac{1}{4} r_4 \zeta_{4,2} \notag\\
 & + \left( \frac{1}{3} r_3 + \frac{1}{4} r_4 \right)
 r_3 r_4 \left( \frac{4}{3} \zeta_{3,1}^2 - 2 \zeta_{3,1} \zeta_{4,1} + \frac{3}{4} \zeta_{4,1}^2 \right).
\label{secondthreefour}
\end{align}
In the $\rho_3$ dominance $r_3 \to 1$, $r_4 \to 0$, while in the  $\rho_4$ dominance 
$r_3 \to 0$, $r_4 \to 1$.
In both $\rho_3$ dominance and $\rho_4$ dominance, since $r_3 r_4 \to 0$, the third term of
the second order Bardeen parameter $D(\rho)^2 \ln{a}$ in (\ref {secondthreefour}) and $D(\rho) r_i$
for $i=3,4$ in (\ref {rparameterderivative}) are negligibly small.
Therefore by operating $D(\rho)$ repeatedly to the above $D(\rho)^2 \ln{a}$ with use of 
(\ref {rparameterderivative}), for an arbitrary natural number $k$ we can obtain 
\begin{equation}
  D(\rho)^k \ln{a} \to \frac{1}{3} \zeta_{3,k}
\end{equation}
for $\rho_3$ dominance, and  
\begin{equation}
  D(\rho)^k \ln{a} \to \frac{1}{4} \zeta_{4,k}
\end{equation}
for $\rho_4$ dominance, respectively. 
The system where the $3$ fluid and the $4$ fluid coexist is often used in the curvaton scenario
\cite{Lyth2002} \cite{Ichikawa2008} \cite{Sasaki2006} \cite{Assadullahi2007} \cite{Suyama2011}.
First the $4$ fluid governs the cosmic energy and the Bardeen parameter, while last the $3$ fluid
governs the cosmic energy and the Bardeen parameter.
Since the $3$ fluid does not contribute to the adiabatic perturbation represented by the Bardeen 
parameter, this phenomenon can be interpreted as the conversion of the entropy perturbation into the 
adiabatic perturbation.

Until now, under the certain assumption which from now on we will call the assumption $NED$ 
(Not Extremely Different), we have stated that the energetically dominant component makes the 
dominant contribution to the Bardeen parameters $D(\rho)^k \ln{a}$ where $k$ is the natural number
and we call this fact the property of oblivion.
By the assumption $NED$, we mean that the sizes of the initial perturbations of many components 
are not extremely different.
Although the equations (\ref{firstsecondfour}), (\ref{secondthreefour}) were derived without putting 
any assumptions, by using the notations of (\ref{firstsecondfour}), (\ref{secondthreefour}),
the assumption $NED$ can be written as $10^{-2} \le I_{NED} \le 10^2$, for instance,
where $I_{NED} := |\zeta_{3,1} / \zeta_{4,1}|$, that is the ratio of the initial perturbations $I_{NED}$ does not 
become extremely large or small.
We will investigate in which case the assumption $NED$ holds or not, using the equations 
(\ref{firstsecondfour}), (\ref{secondthreefour}) with seeing also the observational results.
We will consider the following simple scenario.
The inflationary expansion caused by the field $\phi$ is required to satisfy $N = \kappa^2 \phi(0)^2 \sim 10^2$
in order to guarantee the present observable homogeneous universe, and the $\phi$ field is transformed into
the $4$ fluid.
In addition, the $3$ fluid by the curvaton $\chi$ field oscillation exists.
For simplicity, we assume that the orders of the fields values $\phi (0)$, $\chi(0)$ do not change largely 
between the initial time of the slow rolling phase and the initial time of the fluids phase by $\rho_3$, $\rho_4$.
We write the initial values of the fields $\phi (0)$, $\chi(0)$ as
\begin{equation}
  \phi (0) \sim \frac{1}{\kappa} 10, \quad \chi(0) \sim \frac{1}{\kappa} 10^{- l},
\end{equation}
where $l$ is the nonnegative real number, where by $A \sim B$, we mean $|A/B| + |B/A| \le 0(1)$.
We can derive
\begin{align}
  \zeta_{3,1} &\sim \frac{D \chi(0)}{\chi(0)} + \kappa^2 \phi(0) D \phi(0),\\
  \zeta_{4,1} &\sim \kappa^2 \phi(0) D \phi(0),
\end{align}
where we assume that the base of the derivative operator $D$ is properly chosen, for example, the scale 
factor $a(0)$, since we are only interested in the order estimates.
Since from the equations (\ref{firstsecondfour}), (\ref{secondthreefour}), we can see
\begin{align}
  N_{\phi} &\sim \kappa^2 \phi(0),\label{inflatonpotential1}\\
  N_{\chi} &\sim r_3 \frac{1}{\chi(0)},
\end{align}
and
\begin{align}
 N_{\phi \phi} &\sim \kappa^2 + r_3 r_4 \kappa^4 \phi(0)^2,\label{inflatonpotential2}\\
 N_{\phi \chi} &\sim r_3  r_4 \frac{\kappa^2 \phi(0)}{\chi(0)},\\
 N_{\chi \chi} &\sim r_3  \frac{1}{\chi(0)^2},
\end{align}
we can obtain
\begin{align}
  N_a N^a &\sim \kappa^4 \phi(0)^2 + r^2_3 \frac{1}{\chi(0)^2},\\
  N_{ab} N^a N^b &\sim \kappa^6 \phi(0)^2 + r_3 r_4 \kappa^8 \phi(0)^4 
  + r^2_3 r_4 \kappa^4 \frac{\phi(0)^2}{\chi(0)^2} +r^3_3 \frac{1}{\chi(0)^4}.
\end{align}
From the Planck observation \cite{Planck},  the scalar to tensor ratio ${S^2}/ {T^2}$ and 
the non Gaussianity $f_{NL} $ defined by
\begin{equation}
 \frac{S^2}{T^2} := \frac{1}{\kappa^2}  N_a N^a =: 10^{2 \alpha},\quad
 f_{NL} := \frac{ N_{ab} N^a N^b }{(N_c N^c)^2}
\end{equation}
are constrained as
\begin{equation}
  \alpha \ge 0.5, \quad |f_{NL}| \le O(1).
\label{Planck}
\end{equation}
Many inflationary theorists expect sufficiently small $\alpha$, that is they expect that the gravitational wave contribution 
to the cosmic microwave radiation fluctuation will be able to be detected in the near future.
The index of the assumption $NED$ can be written as
\begin{equation}
 I_{NED} := \left|\frac{\zeta_{3,1}}{\zeta_{4,1}} \right| \sim 10^{l-1} +1. \label{NEDindex}
\end{equation}
While in the $\rho_4$ dominance, by writing $r_3 = 10^{-m}$, the observables are estimated as
\begin{align}
 \frac{S^2}{T^2} &= 10^{2} + 10^{2l-2m} \label{observables1}\\
 f_{NL} &= \frac{10^2 + 10^{4-m}+10^{2+2l-2m} +10^{4l-3m}}{(10^2+10^{2l-2m})^2},
\end{align}
in the $\rho_3$ dominance, by writing $r_4 = 10^{-p}$, the observables are estimated as
\begin{align}
 \frac{S^2}{T^2} &= 10^{2} + 10^{2l} \\
 f_{NL} &= \frac{10^2 + 10^{4-p}+ 10^{2+2l-p}+ 10^{4l}}{(10^2+10^{2l})^2}. \label{observables4}
\end{align}
Since by using the observational results (\ref{Planck}) to the theoretical expressions
(\ref{observables1}) -(\ref{observables4}), we can derive the constraints to the power indexes,
we can estimate the index of the assumption $NED$; $I_{NED}$ in (\ref{NEDindex}).
We consider the $\rho_4$ dominance.
When $m \le l-1$, we get $l=m+\alpha$, $\alpha \ge 1$ and $4 -3 \alpha \le l \le \alpha$.
Then we obtain
\begin{equation}
 I_{NED} \le 10^{\alpha-1}+1,
\end{equation}
implying that the assumption $NED$ is satisfied, 
as long as the gravitational wave contribution to the CMB (Cosmic Microwave Background) fluctuation 
is not extremely small.
When $m > l-1$, $\alpha =1$.
By excluding the uninteresting case that the $\rho_3$ contributions to the first order and the second 
order Bardeen parameters are negligibly small, we obtain the interesting case
that $\rho_3$ contributes the second order significantly while it does not contribute the first order as
\begin{equation}
\min{\left[ \max{\left[\frac{1}{2} m+1, m \right]},
\max{\left[\frac{1}{2} m+1, \frac{3}{4} m + \frac{1}{2} \right]} \right]} < l \le \frac{3}{4} m + 1,
\label{breakNED}
\end{equation}
implying that the index $I_{NED}$ can be large.
When the assumption $NED$ is broken, that is $I_{NED} > 10^2$, we get $l >3$ and $m>8/3$.
In such case, we cannot apply the property of oblivion.
But since in this case $r_3$ is extremely small, the method of solving the Bardeen parameter 
by expanding the evolution equations with respect to $\rho_3$ around $\rho_3 =0$ 
\cite{Hamazaki2011} is efficient.
It is noted that in the $\rho_4$ dominance the role of the observation of the non-Gaussianity is 
essentially important.
Next we consider the $\rho_3$ dominance.
When $l \ge 1$, $l = \alpha$.
We obtain
\begin{equation}
 I_{NED} \le 10^{\alpha-1}+1,
\end{equation}
As long as the gravitational wave contribution to the CMB fluctuation is not very small,
the assumption $NED$ is satisfied.
When $l < 1$, $\alpha =1$.
We get
\begin{equation}
 I_{NED} \sim 1,
\end{equation}
implying that the assumption $NED$ is satisfied.
In summary, the case in which the property of oblivion cannot be applied is confined in the equation
(\ref{breakNED}), and in this case the other efficient method \cite{Hamazaki2011} exists.
In the two step curvatons model \cite{Suyama2011}, the dimension of the parameter space becomes high.
But the similar analysis might be possible.

From the reviewer, we were requested that the above equations (\ref{inflatonpotential1}) (\ref{inflatonpotential2})
should be generalized into the more general potential.
Under the slow rolling condition
\begin{equation}
 |\epsilon|, |\eta| \ll 1,
\end{equation}
where the slow roll parameters are defined by
\begin{equation}
 \epsilon := \frac{1}{\kappa^2 U^2} \left( \frac{d U}{d \phi} \right)^2, \quad \quad
 \eta := \frac{1}{\kappa^2 U} \frac{d^2 U}{d \phi^2}, 
\end{equation}
the evolution equation of the scalar field 
\begin{equation}
 \frac{d \phi}{d N} = - \frac{1}{\kappa^2 U} \frac{d U}{d \phi},
\label{slowrollscalarevo}
\end{equation}
can be solved as
\begin{equation}
 N - N(0) = \kappa^2 \int^{\phi(0)}_{\phi} d \phi \left[ U \Big/ \frac{d U}{d \phi} \right],
\label{slowrollscalarsol}
\end{equation}
which was given in the paper \cite{Hamazaki2008.2}, since the first order ordinary differential equation of the type
of the separation of variables is always integrable.
Since by using the slow roll parameters $\epsilon$, $\eta$ under the evolution equation (\ref{slowrollscalarevo})
\begin{equation}
 \frac{\ddot{\phi}}{3 H \dot{\phi}} = \frac{1}{6} \epsilon- \frac{1}{3} \eta,
\end{equation}
can be derived, $\ddot{\phi}$ in the exact evolution equation of the scalar field $\phi$ can be approximately 
discarded. 
By taking the derivatives of (\ref{slowrollscalarsol}), we can obtain the more general expressions than
(\ref{inflatonpotential1}) (\ref{inflatonpotential2}):
\begin{align}
 N_{\phi} &= \kappa^2 \left[ U \Big/ \frac{d U}{d \phi} \right] \Bigg|_{\phi = \phi (0)} 
 = \pm \frac{\kappa}{\sqrt{\epsilon}}, \\
 N_{\phi \phi} &= \kappa^2 \left[ 1 
 - U \frac{d^2 U}{d \phi^2} \Big/ \left( \frac{d U}{d \phi} \right)^2 \right] \Bigg|_{\phi = \phi (0)}
 = \kappa^2 \left(1 - \frac{\eta}{\epsilon} \right).
\end{align}
Therefore the estimations (\ref{inflatonpotential1}) (\ref{inflatonpotential2}) hold in the case that 
the potential is written by the single power of the scalar field $\phi$: 
\begin{equation}
 U = \lambda_n \phi^n,
\end{equation}
where $n$ is the integer and $\lambda_n$ is the constant.

We consider the curvaton scenario in which the inflaton scalar field $\phi$ obeys the potential
defined by
\begin{equation}
 U = U_0 - \frac{\lambda}{4} \phi^4,
\end{equation}
where $\phi$ starts from near the potential maximum, $U_0$ is the constant driving the inflationary expansion,
$\lambda$ is the small coupling constant.
The initial values of the fields $\phi (0)$, $\chi(0)$ are written as
\begin{equation}
  \phi (0) \sim \frac{\eta}{\kappa}  10^{-1}, \quad \chi(0) \sim \frac{1}{\kappa} 10^{- l},
\end{equation}
where by using the non-negative real number $q$
\begin{equation}
 \eta := \frac{\kappa^2 U_0^{1/2}}{\lambda^{1/2}} = 10^q
\end{equation}
and $l$ is the nonnegative real number.

We can derive
\begin{align}
  \zeta_{3,1} &\sim \frac{D \chi(0)}{\chi(0)} + \frac{\eta^2}{\kappa^2} \frac{D \phi(0)}{\phi(0)^3},\\
  \zeta_{4,1} &\sim \frac{\eta^2}{\kappa^2} \frac{D \phi(0)}{\phi(0)^3} ,
\end{align}
where we assume that the base of the derivative operator $D$ is properly chosen, for example, the scale 
factor $a(0)$.
The index of the assumption $NED$ can be written as
\begin{equation}
 I_{NED} := \left|\frac{\zeta_{3,1}}{\zeta_{4,1}} \right| \sim 10^{l+q-3} +1. \label{NEDindexnew}
\end{equation}
While in the $\rho_4$ dominance, by writing $r_3 = 10^{-m}$, the observables are estimated as
\begin{align}
 \frac{S^2}{T^2} &= 10^{6-2q} + 10^{2l-2m} \label{observables1}\\
 f_{NL} &= \frac{10^{10- 4 q} + 10^{12- 4q- m}+10^{7+ l - 3 q- 2 m} +10^{4l-3m}}{(10^{6-2q}+10^{2l-2m})^2},
\end{align}
in the $\rho_3$ dominance, by writing $r_4 = 10^{-p}$, the observables are estimated as
\begin{align}
 \frac{S^2}{T^2} &= 10^{6 - 2 q} + 10^{2l} \\
 f_{NL} &= \frac{10^{10- 4 q} +  10^{12- 4 q - p}+ 10^{7+ l - 3 q - p}+ 10^{4l}}{(10^{6- 2 q}+10^{2l})^2}. 
\end{align}
In the same way as in the above case $N \sim \kappa^2 \phi(0)^2$ by using the observational results 
(\ref{Planck}) to the above theoretical expressions,
we will estimate the index of the assumption $NED$; $I_{NED}$ in (\ref{NEDindexnew}).
We consider the $\rho_4$ dominance.
When $m \le l+q-3$, we get $l=m+\alpha$ and $l \le \alpha$.
Then we obtain
\begin{equation}
 I_{NED} \le 10^{\alpha+q-3}+1,
\end{equation}
implying that the assumption $NED$ is satisfied, 
as long as the gravitational wave contribution to the CMB (Cosmic Microwave Background) fluctuation 
is not extremely small and $q$ is not extremely large.
Since in the same way as in the above case $N \sim \kappa^2 \phi(0)^2$, in the other cases, 
we can estimate $I_{NED}$, we will omit the detail.
Also in the one curvaton model based on the new inflation paradigm, the observational results of 
the scalar to tensor ratio and the non-Gaussianity are useful in order to constrain the index of the assumption $NED$;
$I_{NED}$.

We got to know that unless the assumption $NED$ is satisfied, that is if the amplitude of the scalar field 
oscillation $\chi(0)$ is very small, the energetically 
subdominant component can make the significant contribution of the Bardeen parameter.
Then in this case, is the idea of the property of oblivion meaningless at all?
In relation to the equations (\ref{firstsecondfour}), (\ref{secondthreefour}), as for the system in which 
the $3$ fluid $\rho_3$ and the $4$ fluid $\rho_4$ coexist, without assuming the assumption $NED$,
we can easily derive the following statement:

\paragraph{Proposition}
The finite order Bardeen parameter $D(\rho)^k \ln{a}$ where $k$ is the natural number can be written 
in the form of the polynomial constructed from the products of the initial perturbations $\zeta_{3,l}$ and
$\zeta_{4,m}$, that is $\Pi_{\alpha} \zeta_{3,l_{\alpha}} \Pi_{\beta} \zeta_{4,m_{\beta}}$ 
($\sum_{\alpha} l_{\alpha} + \sum_{\beta} m_{\beta} = k$).
We assume that the index $A$ is one element of the set $\{3, 4\}$ and 
that the index $B$ is the other element of the set $\{3, 4\}$, that is $A \neq B$.
We assume that $\rho_A$ is the energetically dominant component and that $\rho_B$ is the energetically
subdominant component.
For all the coefficients of the products other than $\zeta_{A, k}$, the absolute values of the coefficients 
are suppressed by $C_{l,m} r_B$ where $C_{l,m}$ is the constants or order of unity.

When the assumption $NED$ is satisfied, since the contributions to the Bardeen parameter are governed by
the sizes of the coefficients, the contribution from the subdominant component becomes small.
On the other hand, when the assumption $NED$ is not satisfied, if the initial perturbation $\zeta_{B,p}$
is large enough to overwhelm the smallness of the coefficient suppressed by $r_B$,
the contribution from the energetically subdominant component can be large.
But we will not pursue such possibility \cite{Sasaki2006} \cite{Suyama2011} any further.

Next we consider the relation with the paper \cite{Wands2000} written under the name of the 
separate universe approach.
In the separate universe approach, the perturbation in the long wavelength limit is interpreted as
the difference between the values of the physical quantity at two points separated by the super
horizon scales around which the physical quantity obeys the evolution equation and the constraint
condition of the exactly homogeneous universe.
We reinterpret their statement by our notation.
For the arbitrary scalar like objects $S_1$, $S_2$ and for an arbitrary natural number $m$, 
as the $m$-th order perturbation of $S_2$ in the time slice 
where $S_1$ is not perturbed, instead of using $D(S_1)^m S_2$, the authors of  the paper 
\cite{Wands2000} use $D_i (S_1)^m S_2$, where for an arbitrary scalar like object $A$,
\begin{equation}
 D_i (S_1) A:= 
\left[ \frac{d}{d x^i}  - \left( \frac{d S_1}{d x^i} \Bigg/ \frac{d S_1}{d t}\right) \frac{d}{d t} \right] A,
\end{equation}
where $i=1,2,3$ and we use $d$ instead of $\partial$ for notational simplicity.
$D_i (S_1)^m S_2$ is obtained by replacing the perturbation parameter $\lambda$ with the spatial
coordinate $x^i$ ($i=1,2,3$) in the definition of $D(S_1)^m S_2$.
$D_i (S_1)^m S_2$ is the scalar like object, while $D(S_1)^m S_2$ is the background/scalar like object
\cite{Hamazaki2011}.
In the zeroth order of the gradient expansion, the time components of the evolution equations and 
the Hamiltonian constraint have the same form as those of the exactly homogeneous universe 
and all the evolution equations and all the constraint equations are written as the polynomials of 
the scalar like objects \cite{Hamazaki2011}.
In order to obtain the perturbation equations,
while we operate $D(S_1)^m$ on those equations, the authors of the paper \cite{Wands2000}
operate $D_i (S_1)^m$ on them.
Therefore the both are equivalent under the correspondence 
$D(S_1)^m S_2 \leftrightarrow D_i (S_1)^m S_2$.

In order to derive the part of the main results of the paper \cite{Wands2000}, we consider the many 
fluids system whose component fluids have the definite equations of state $P_k = P_k (\rho_k)$ where
$P_k$, $\rho_k$ are the pressure, the energy density of the fluid $k=1,2, \cdots$, respectively.
As noted in the first part of the section \ref{mainchangebase},  for arbitrary scalar like objects $A$,
$S$, we interpret the evolution operator $d / d S$ as
\begin{equation}
 \frac{d A}{d S} := \frac{d A}{d t} \Big/ \frac{d S}{d t}.
\end{equation}
First $D(\rho_k) \ln{a}$ ($k=1,2,\cdots$) is conserved since
\begin{equation}
 \frac{d}{d \rho_k} D(\rho_k) \ln{a} = D(\rho_k) \frac{d}{d \rho_k} \ln{a} 
= D(\rho_k) \left[ - \frac{1}{3} \frac{1}{\rho_k + P_k (\rho_k)}  \right] =0,
\end{equation}
where in the first equality we use the theorem \ref{mainchangebase}.$1$.
In the same way as in the above, as for $D(\rho) \ln{a}$, we obtain the famous formula
\begin{equation}
 \frac{d}{d \rho} D(\rho) \ln{a} = \frac{1}{3} \frac{1}{(\rho + P)^2} D(\rho) P, 
\end{equation}
where $D(\rho) P$ is the entropy perturbation variable \cite{Hamazaki2011} defined as 
the perturbation variable which vanishes for the adiabatic growing mode explained later.
From this formula, it can be said that
in the case where the pressure $P$ is the definite function of the energy density $\rho$,
such as the case the universe consists of a single fluid, the Bardeen parameters in full order 
are conserved.
This famous fact is the central theme of the paper \cite{Lyth2005}.
By this formula, we can also say the following fact:
If $ D(\rho) \rho_k$ for all $k$ vanish at some time, for all the subsequent time all the Bardeen 
parameters  $D(\rho)^m \ln{a}$ for all natural number $m$ are conserved,
since the following equations 
\begin{equation}
 \frac{d}{d \rho} D(\rho) \rho_k = \sum_j A_{kj} \cdot D(\rho) \rho_j, \quad
 D(\rho) P = \sum_j B_j \cdot D(\rho) \rho_j,
\end{equation}
hold for some analytic functions of the energy densities of the component fluids $\rho_k$; $A_{kj}$, $B_j$
($k,j =1,2,\cdots$).
Such solution is called the adiabatic growing mode and is constructed in the following way \cite{Kodama1998}:
If $a$, $S$ where $a$ is the scale factor and $S$ is all the other scalar like objects, is the solution
of the zeroth order of the evolution equations and the constraint equations, 
$a (1+ \lambda_{\ast})$, $S$ where $\lambda_{\ast}$ is the constant is also the solution, since
the scale factor $a$ always appears in the form of the fraction such as $H:=\dot{a} / a$ in the zeroth order
of those equations.
Then we differentiate such expression with respect to $\lambda_{\ast}$.

As for the many fluid system, we, the paper \cite{Sasaki2006} and the paper \cite{Wands2000} use only the 
equations of the conservation of the energy momentum as
\begin{equation}
 a \frac{d}{d a} \rho_k = -3 (\rho_k + P_k).
\end{equation}
So our results and the results of the papers \cite{Sasaki2006} \cite{Wands2000} hold in all the gravitational 
theories keeping the general covariance and the conservation of the energy momentum.
On the other hand, since we use the Hamiltonian constraint in the analysis of the evolution of the slow rolling 
multiple scalar fields in the section \ref{mainchangebase}, this consideration can only be applied 
to the general theory of relativity, that is, the Einstein theory of gravity.

\addtolength{\baselineskip}{-3mm}

%T1>Biblipgraphy


\begin{thebibliography}{10}

\bibitem{Sato1981}
{Sato, K.}, 
"First-order phase transition of a vacuum and the expansion of the Universe",
Mon.Not.Roy.Astro.Soc.{\bf 195}, 467-479 (1981).

\bibitem{Guth.1981}
{Guth, A.H.}, 
"Inflationary universe: A possible solution to the horizon and flatness problems",
Phys. Rev. D {\bf 23}, 347 (1981).

\bibitem{Bardeen.J1980}
{Bardeen, J.M.}, 
"Gauge Invariant Cosmological Perturbations",
Phys. Rev. D {\bf 22}, 1882-1905 (1980).

%entropic perturbation
\bibitem{Kodama1984}
{Kodama, H. } and { Sasaki, M.}, 
"Cosmological Perturbation Theory",  
Prog. Theor. Phys. Suppl. {\bf 78}, 1-166
  (1984).

\bibitem{Kodama1987}
{Kodama, H. } and { Sasaki, M.}, 
"Evolution of Isocurvature Perturbations. 2. Radiation Dust Universe",
Int. J. Mod. Phys. {\bf A2}, 491
  (1987).

\bibitem{Mukhanov1992}
{Mukhanov, V.F.}, {Feldman, H.A.} and {Brandenberger, R.H.},
"Theory of cosmological perturbations",
Phys. Rep. {\bf 215}, 203-333 (1992).

\bibitem{Gordon2001}
{ Gordon, C.}, {Wands, D.}, {Bassett, B.A.} and {Maartens, R.}, 
"Adiabatic and entropy perturbations from inflation",
 Phys. Rev D{\bf 63} 023506 (2001).

%long wavelength linear my work
\bibitem{Kodama1996}
{Kodama, H. } and { Hamazaki, T.},
"Evolution of cosmological perturbations in a stage dominated by an oscillatory scalar field",
 Prog. Theor. Phys. {\bf 96},949-970 (1996).

\bibitem{Hamazaki1996}
{ Hamazaki, T.} and {Kodama, H.},
" Evolution of cosmological perturbations during reheating", 
 Prog. Theor. Phys. {\bf 96},1123-1146
  (1996).

\bibitem{Kodama1998}
{Kodama, H. } and { Hamazaki, T.}, 
"Evolution of cosmological perturbations in the long wavelength limit", 
Phys. Rev. D{\bf 57}, 7177-7185
  (1998).

\bibitem{Hamazaki2002}
{ Hamazaki, T.}, 
"Evolution of cosmological perturbations in the universe dominated by multiple scalar fields", 
Phys. Rev. D {\bf 66}, 023529 (2002).

\bibitem{Hamazaki2004}
{ Hamazaki, T.}, 
"Evolution of cosmological perturbations in the universe dominated by resonant scalar fields", 
Nucl. Phys. B {\bf 698},335-385 (2004).

\bibitem{Hamazaki2008}
{ Hamazaki, T.}, 
"Long wavelength limit of evolution of cosmological perturbations in the universe where scalar fields and fluids coexist",  
Nucl. Phys. B {\bf 791},20-59 (2008).

\bibitem{Hamazaki2008.2}
{ Hamazaki, T.},
" Long wavelength limit of evolution of nonlinear cosmological perturbations", 
 Phys. Rev. D {\bf 78}, 103513 (2008).

\bibitem{Hamazaki2011} 
{ Hamazaki, T.},
"Manifestly gauge invariant theory of the nonlinear cosmological perturbations in the leading order of the gradient expansion",  
 Phys. Rev. D {\bf 84}, 023502 (2011).
 
%long wavelength limit linear

\bibitem{Polarski1992} 
{Polarski, D} and {Starobinski, A}
 "Spectra of perturbations produced by double inflation with an intermediate matter dominated stage"
 Nucl.Phys. B{\bf385} (1992) 623-650  

\bibitem{Taruya1998}
{Taruya, A. } and { Nambu, Y.}, 
"Cosmological perturbation with two scalar fields in reheating after inflation",
Phys. Lett. B428 37-43
  (1998). 

\bibitem{Sasaki1998}
{Sasaki, M. } and { Tanaka, T }, 
"Superhorizon scale dynamics of multiscalar inflation", 
Prog. Theor. Phys. {\bf 99},763-782
  (1998). 

\bibitem{Wands2000}
{Wands, D.}, {Malik, K.A.}, {Lyth, D.H.} and {Liddle, A.R.}
"A New approach to the evolution of cosmological perturbations on large scales",
Phys. Rev. D{\bf 62}, 043527 (2000).

\bibitem{Lyth2005}
 {Lyth, D.H.}, { Malik, K.A.} and {Sasaki, M.},
"A General proof of the conservation of the curvature perturbation",
 JCAP. 0505, 004 (2005).

%\bibitem{Nambu2006}
%{Nambu, Y. } and { Araki, Y.}, Class. Quant. Grav. 23,511 (2006).

\bibitem{Malik2009}
{Malik, K.A.} and {Wands, D.},
"Cosmological perturbations",
Phys. Rept. {\bf 475}, 1-51 (2009).

\bibitem{Nakamura2006}
{Nakamura, K.},
"Gauge-invariant formulation of the second-order cosmological perturbations",
Phys. Rev. D{\bf 74}, 101301 (2006).

%multistep reheating

\bibitem{Elliston2014}
{Elliston, J.}, {Orani, S.} and {Mulryne, D.J.},
"General analytic predictions of two-field inflation and perturbative reheating",
Phys. Rev. D{\bf 89}, 103532 (2014).

\bibitem{Meyers2014}
{Meyers, J.} and {Tarrant, E.R.M.},
"Perturbative Reheating After Multiple-Field Inflation: The Impact on Primordial Observables",
Phys. Rev. D{\bf 89}, 063535 (2014).

\bibitem{Leung2012}
{Leung, G.}, {Tarrant, E.R.M.}, {Byrnes,C.T.} and {Copeland, E.J.},
"Reheating, multifield inflation and the fate of the primordial observables",
JCAP.{\bf 1209}, 008 (2012).

%junction
\bibitem{Israel1966}
 {Israel, W.} ,
"Singular hypersurfaces and thin shells in general relativity"
Nuovo. Cim. {\bf 44B}, 1–14 (1966).

\bibitem{Deruelle1995}
 {Deruelle, N.} and {Mukhanov, V.F.},
"On matching conditions for cosmological perturbations",
Phys. Rev. D{\bf 52}, 5549-5555 (1995).

\bibitem{Martin1998}
 {Martin, J.} and {Schwarz, D.J.},
"The Influence of cosmological transitions on the evolution of density perturbations",
Phys. Rev. D{\bf 57}, 3302-3316 (1998).

\bibitem{Copeland2007}
{Copeland, E.J.} and {Wands, D.},
"Cosmological matching conditions",
 JCAP{\bf 0706}:014 (2007).

\bibitem{Mukohyama2000}
{ Mukohyama, S.},
"Perturbation of junction condition and doubly gauge invariant variables",  
 Class. Quant. Grav. {\bf 17}:4777-4798 (2000).
 
%modulated reheating
\bibitem{Sasaki1991}
{Sasaki, Y.} and {Yokoyama, J.},
"Initial condition for the minimal isocurvature scenario",
Phys. Rev. D{\bf 44}, 970-979 (1991).

\bibitem{Dvali2004}
{Dvali, G.}, {Gruzinov, A.} and {Zaldarriga, M.},
"A new mechanism for generating density perturbations from inflation",
 Phys. Rev. D {\bf 69}, 023505
 (2004).
 
%curvaton
\bibitem{Lyth2002}
{D. H. Lyth} and {D. Wands},
"Generating the curvature perturbation without an inflaton",
 Phys. Lett. B {\bf 524}, 5–14 (2002).

\bibitem{Ichikawa2008}
 {Ichikawa, K.}, {Suyama, T}, {Takahashi, T} and {Yamaguchi, M.},
"Non-Gaussianity, Spectral Index and Tensor Modes in Mixed Inflaton and Curvaton Models",
Phys. Rev. D{\bf 78}, 023513 (2008).

\bibitem{Sasaki2006}
{M.Sasaki}, {J.Valiviita} and {D.Wands}, 
"Non-Gaussianity of the primordial perturbation in the curvaton model",
Phys. Rev. D {\bf 74}, 103003
 (2006).

\bibitem{Assadullahi2007}
{H.Assadullahi}, {J.Valiviita} and {D.Wands},
"Primordial non-Gaussianity from two curvaton decays",
 Phys. Rev. D {\bf 76}, 103003
 (2007).

\bibitem{Suyama2011}
{T.Suyama} and {J.Yokoyama}, 
"Temporal enhancement of super-horizon curvature perturbations from decays of two curvatons and its
cosmological consequences",
Phys. Rev. D {\bf 84}, 083511
 (2011).

%CMB nonlinearity
\bibitem{Komatsu2001}
 {Komatsu, E.} and {Spergel, D.N.},
"Acoustic signatures in the primary microwave background bispectrum",
Phys. Rev. D{\bf 63}, 063002 (2001).

\bibitem{Nakamura2014}
{Nakamura, K.},
"Recursive structure in the definitions of gauge-invariant variables for any order
perturbations",
Class.Quant.Grav. 31 (2014) 135013

\bibitem{Giesel2010}
{Giesel,K} ,{Hofmann,S} ,{Thiemann,T} and {Winkler,O}
"Manifestly gauge-invariant general relativistic perturbation theory: I. Foundations"
Class.Quant.Grav.27 (2010) 055005

\bibitem{Kodama1991}
{Kodama, H}  "Relativistic Cosmology" (1991) Parity Physics Course, Maruzen
(Japanese)

\bibitem{Planck}
Planck Collaboration
"Planck 2015 results. XVII. Constraints on primordial non-Gaussianity"arXiv:1502.01592
"Planck 2015 results. XX. Constraints on inflation"arXiv:1502.02114 and others

\end{thebibliography}
\end{document}